\documentclass[amsmath,amssymb, aps, pra,reprint,footinbib,floatfix,superscriptaddress]{revtex4-2}

\usepackage{amsthm}
\usepackage{amsmath,bm}
\usepackage{amssymb}
\usepackage{amsfonts}
\usepackage{graphicx}
\usepackage{txfonts}
\usepackage[squaren]{SIunits}
\usepackage[dvipsnames]{xcolor}

\usepackage{soul}
\usepackage{float}

\usepackage[colorlinks=true,linkcolor=blue,citecolor=blue,urlcolor=blue]{hyperref}

\usepackage{braket}

\newcommand{\diff}{\mathrm{d}}
\newcommand{\da}{\dagger}
\newcommand{\sinc}{\mathrm{sinc}}
\newcommand{\Op}[1]{\mathsf{#1}}
\renewcommand{\vec}[1]{\boldsymbol{#1}}
\newcommand{\vq}{\vec{q}}

\newcommand{\vu}{\vec{u}}
\renewcommand{\vr}{\vec{r}}

\newcommand{\ve}{\mathbf{e}}

\newcommand{\ovr}{\vec{\Op{r}}}

\newcommand{\oL}{\Op{L}}
\newcommand{\oa}{\Op{a}}
\newcommand{\ox}{\Op{X}}
\newcommand{\op}{\Op{P}}

\newcommand{\cD}{\mathcal{D}}
\newcommand{\cL}{\mathcal{L}}
\newcommand{\cI}{\mathcal{I}}

\newcommand{\eg}{\textit{e.g.} }
\newcommand{\ie}{\textit{i.e.} }

\usepackage{multirow}

\begin{document}

\title{Macroscopic quantum test with bulk acoustic wave resonators}

\author{Bj\"orn Schrinski}
\affiliation{Center for Hybrid Quantum Networks (Hy-Q), Niels Bohr Institute, University of Copenhagen, Blegdamsvej 17, DK-2100 Copenhagen, Denmark}

\author{Yu Yang}
\affiliation{Department of Physics, ETH Z\"urich, 8093 Z\"urich, Switzerland} 
\author{Uwe von L\"upke}
\affiliation{Department of Physics, ETH Z\"urich, 8093 Z\"urich, Switzerland} 
\author{Marius Bild}
\affiliation{Department of Physics, ETH Z\"urich, 8093 Z\"urich, Switzerland} 
\author{Yiwen~Chu}
\affiliation{Department of Physics, ETH Z\"urich, 8093 Z\"urich, Switzerland} 

\author{Klaus Hornberger}
\affiliation{University of Duisburg-Essen, Faculty of Physics, Lotharstraße 1, 47048 Duisburg, Germany} 

\author{Stefan Nimmrichter}
\affiliation{Naturwissenschaftlich-Technische Fakult\"at, Universit\"at Siegen, Walter-Flex-Straße 3, 57068 Siegen, Germany} 

\author{Matteo Fadel}
\email{fadelm@phys.ethz.ch}
\affiliation{Department of Physics, ETH Z\"urich, 8093 Z\"urich, Switzerland} 

\date{\today}

\begin{abstract}
    Recently, solid-state mechanical resonators have become a platform for demonstrating non-classical behavior of systems involving a truly macroscopic number of particles.
    Here, we perform the most macroscopic quantum test in a mechanical resonator to date, which probes the validity of quantum mechanics by ruling out a classical description at the microgram mass scale. This is done by a direct measurement of the Wigner function of a high-overtone bulk acoustic wave resonator mode, monitoring the gradual decay of negativities over tens of microseconds. While the obtained macroscopicity of $\mu= 11.3$ is on par with state-of-the-art atom interferometers, future improvements of mode geometry and coherence times could test the quantum superposition principle at unprecedented scales and also place more stringent bounds on spontaneous collapse models.
\end{abstract}

\maketitle

Understanding the quantum-classical transition is one of the main challenges of modern physics. Is the Schr\"odinger equation valid all the way from the microscopic to the macroscopic world, with quantum effects increasingly hard to observe due to environmental decoherence \cite{schlosshauer2019quantum}? Or do the laws of quantum mechanics break down at some point, so as to reinstate \textit{macrorealism} in our everyday life \cite{leggett2002testing}? This question is not  philosophical, as it can be tackled by demonstrating genuine quantum effects in ever more macroscopic systems.

Single atoms have been delocalized over macroscopic length and time scales, reaching metres and seconds in state-of-the-art experiments \cite{kovachy2015quantum,abend2016atom,asenbaum2017phase,xu2019probing}. On the other hand, molecule interferometry has pushed the mass boundaries by confirming the wave nature for compounds of more than $10^4$ atomic mass units 
\cite{eibenberger2013matter,fein2019quantum}. Future Earth- and space-based experiments with freely falling or levitated nanoparticles aim at even higher mass and time scales \cite{stickler2018probing,kaltenbaek2022maqro}, but in such schemes it is a single rigid-body degree of freedom that is placed into a superposition.

This is to be contrasted with growing efforts in electro- and opto-mechanics toward demonstrating quantum states in the elastic deformation of condensed matter systems of substantially higher mass. Recent experimental milestones include the successful preparation of nonclassical states of surface and bulk acoustic modes, demonstrated through the observation of Wigner function negativities \cite{Chu2018,Satzinger18,Wollack22,von2022parity,bild2022}. Such nonclassical states of motion involve a truly macroscopic number of atoms, on the order of $10^{16}$, vibrating in unison. At the same time, each atom is delocalized only over a tiny fraction of the size of an atomic nucleus, which makes unclear how they compare to previous tests of the quantum-classical transition.

Using an objective measure for the degree of macroscopicity of nonclassical states \cite{nimmrichter2013macroscopicity}, we report here the most macroscopic quantum test performed with a solid-state mechanical resonator. 
The macroscopicity value obtained in our experiment is surprisingly large, motivating us to optimize the design specifications for future electromechanical experiments to become competitive even with matter-wave superposition tests.

\begin{figure}[t]
    \centering
    \includegraphics[width=0.85\columnwidth]{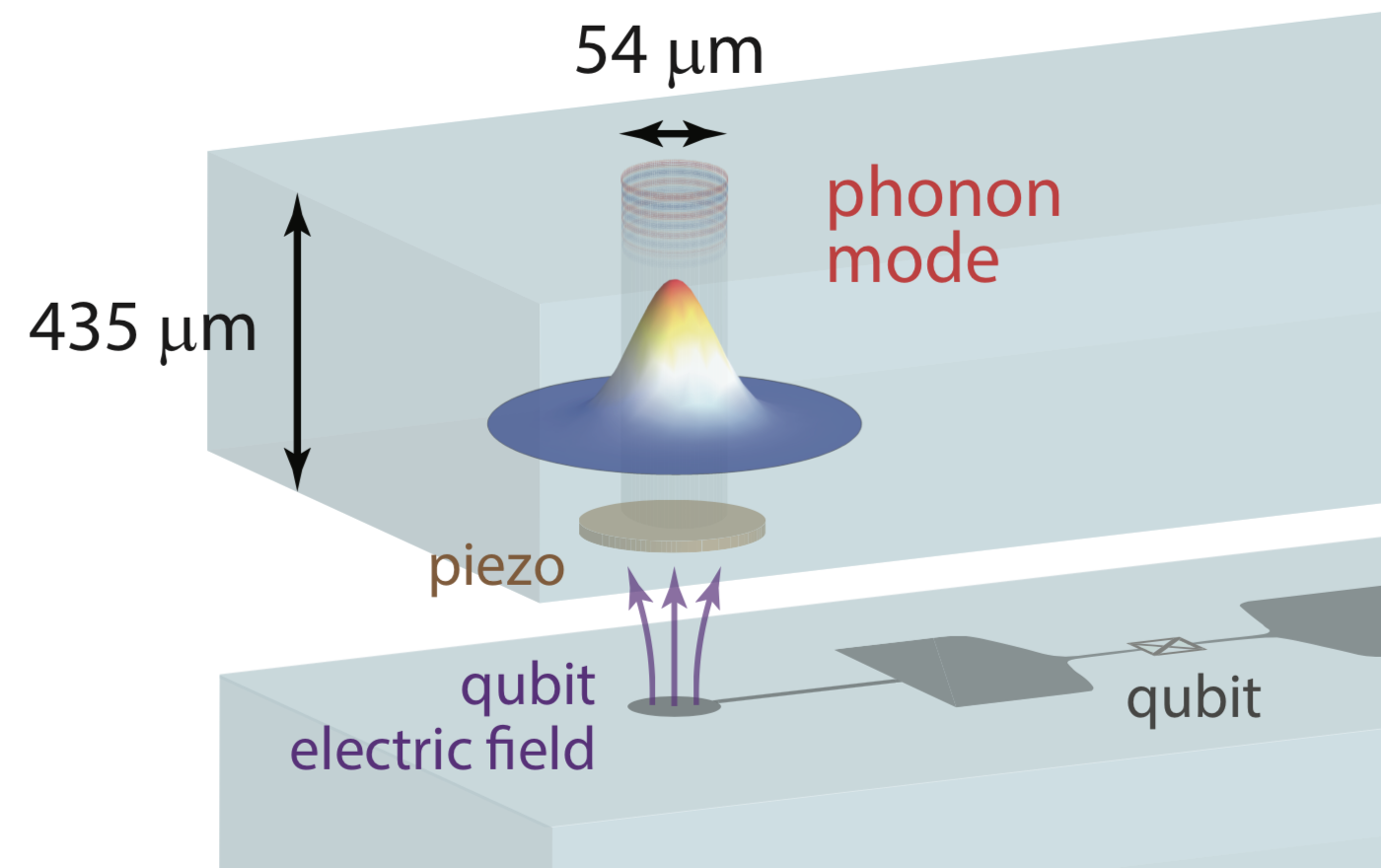}
    \caption{Illustration of the HBAR device used in this work (not to scale, see \cite{von2022parity} for details). A superconducting qubit couples to a mechanical acoustic mode localized in the bulk of a sapphire crystal through a piezoelectric material. This allows preparing and measuring the quantum state of the phonon mode.}
    \label{fig:sketch}
\end{figure}

To justify the macroscopicity measure used, note that
any genuine quantum test probes the validity of quantum mechanics against the hypothesis of macrorealism: that a modification of quantum theory destroys coherence above a certain size or mass scale, rendering superposition phenomena unobservable in practice.
Hence, the degree to which a quantum experiment falsifies a broad, generic class of minimally invasive, macrorealistic modifications of quantum mechanics gives rise to an objective benchmark \cite{nimmrichter2013macroscopicity}: Using the measurement data to rule out a range of values of the coherence time parameter $\tau_e$ quantifying the macrorealist modification, an experiment is  justifiably the more macroscopic the greater the $\tau_e$-values falsified through Bayesian inference \cite{schrinski2019macroscopicity}. The macroscopicity measure $\mu$ is then determined by the logarithm of the greatest ruled out $\tau_e$-value \footnote{Note that the macrorealist modification hypothesis can be probed also in experiments that to do not involve quantum superpositons. Only observations of genuine quantum signatures can therefore be assigned a benchmark value $\mu$ \cite{schrinski2019macroscopicity}. This requires the interrogated state not to have an underlying classical description, excluding all oscillator states with positive Wigner function.}. To date, the highest macroscopicity $\mu = 14 $ has been achieved in molecule interferometry \cite{fein2019quantum}.

In our experiment, we observe Wigner function negativities for up to $40\,\mu$s in an bulk acoustic vibration mode of a sapphire crystal with an effective mass of $1\,\mu $g, which is prepared in a single-phonon state and in a superposition of Fock states.
This quantum signature is detected by measuring the displaced parity  \cite{royer1977wigner} of the quantum state using a superconducting qubit, see Fig.~1. Our analysis shows that it can be associated with a large macroscopicity $\mu = 11.3$, despite the sub-nuclear delocalization of the atoms involved. Based on this, we explain how future experiments with bulk acoustic resonators may realize the most macroscopic quantum test.

\begin{figure*}
  \centering
\includegraphics[width=\textwidth]{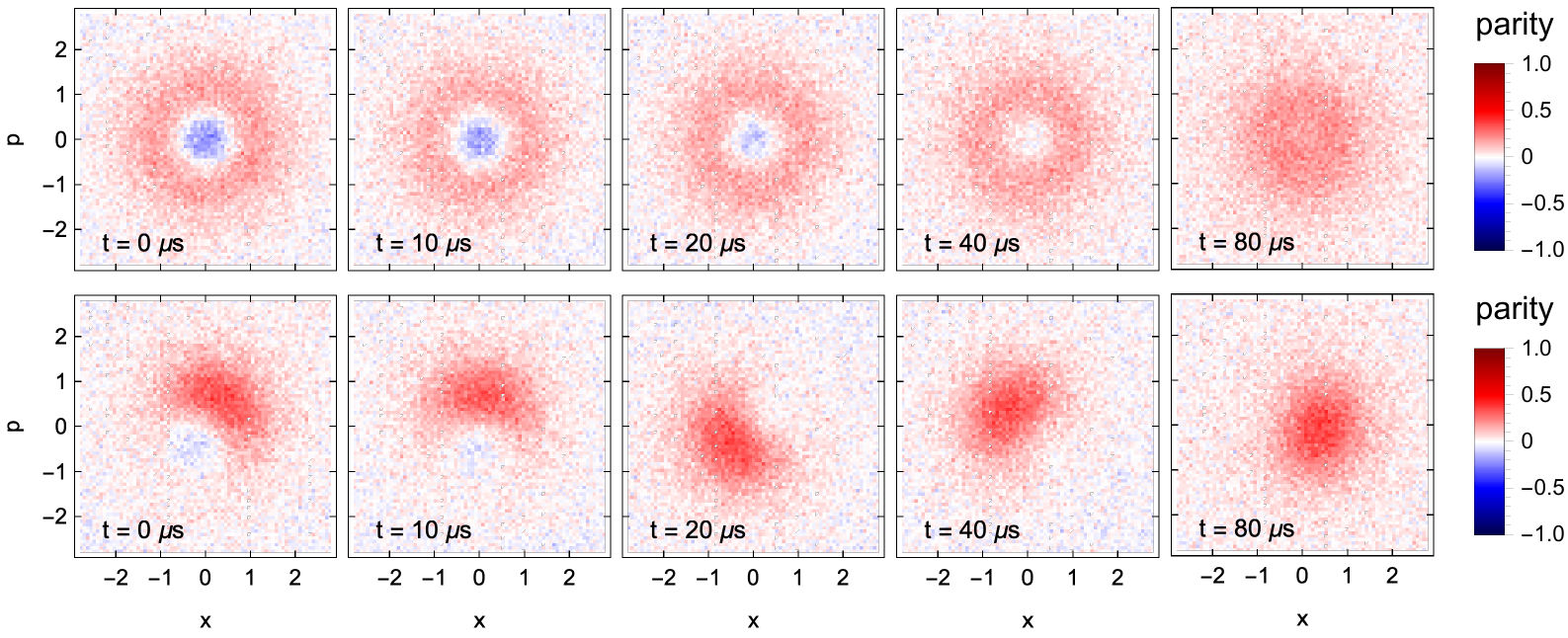}
  \caption{%
  Measured Wigner functions of a bulk acoustic mode initially prepared in the Fock state $\ket{1}$ (top row) and in the superposition state $(\ket{0}+\ket{1})/\sqrt{2}$ (bottom row). Both states are initially non-classical, as indicated by the phase space regions with negative quasi-probability (blue). As time evolves (left to right), both states approach the ground state $\ket{0}$ due to relaxation.
  Comparing the measured data to the theoretical model Eq.~\eqref{eq:TimeWigner} allows us to extract the maximal compatible diffusion rate $\Gamma$ through Bayesian inference, yielding the macroscopicity $\mu$ for the considered experiment.
  }\label{fig2}
\end{figure*}

\vspace{2mm}
\emph{Macroscopicity of bulk resonator modes.}-- %
To evaluate the macroscopicity of a demonstrated quantum effect, one should consider the class of macrorealistic models possibly falsified by the experiment. Keeping those models minimally invasive in that they preserve linearity, the semigroup nature of the time evolution, invariance under the chosen non-relativistic reference frame, and other basic consistency principles \cite{nimmrichter2013macroscopicity}, their observable impact can be formulated as a Lindblad generator $\cL \rho$ modifying the quantum evolution of the system state $\rho$. For a large compound of $N\gg 1$ atomic constituents with masses $m_1,\ldots,m_N$, the generator reduces to \cite{nimmrichter2013macroscopicity}
\begin{equation}\label{eq:CLT}
\cL \rho = \frac{1}{\tau_e}\int \diff^3q\,g(q,\sigma_q)\cD[\oL(\vq)]\rho \;,
\end{equation} 
with $\cD[\oL]\rho = \oL\rho\oL^\da - \{\oL^\da \oL,\rho \}/2$. The strength of the modification is specified by the coherence time parameter $\tau_e$ and the standard deviation $\sigma_q$ of the isotropic Gaussian momentum distribution $g(q,\sigma_q)= e^{-q^2/2\sigma_q^2}/(2\pi\sigma_q^2)^{3/2}$. The Lindblad operators are mass-weighted sums of single-particle momentum kicks by $\vq$, $\oL(\vq)=\sum_{n=1}^N (m_n/m_e) e^{i\vq\cdot\ovr_n/\hbar}$, with $\ovr_n$ the particle position operators and $m_e$ a  reference mass chosen to be that of an electron.  

We note that the theory of continuous spontaneous collapse (CSL) can be described by \eqref{eq:CLT}, but it is usually parameterized in terms of a rate per atomic mass unit $\lambda_{\mathrm{CSL}} = (1\,\mathrm{u}/m_e)^2/\tau_e$, and a localisation length $r_{\mathrm{CSL}} = \hbar/\sqrt{2}\sigma_q$ \cite{bassi2013models}. 
The strongest bounds on these CSL parameters are currently achieved by so-called non-interferometric tests \cite{CarlessoNat22}, where one measures the noise experienced by a classical system by exploiting the fact that macrorealistic modifications are accompanied by a heating effect.
However, a genuine quantum signal, such as interference fringes or a negative Wigner function, is required for a CSL test to also probe the validity of quantum mechanics in the system.

In the following, we consider a collective acoustic excitation in the bulk condensed matter,  
taking the form 
\begin{equation}\label{eq:LG00mode}
    \vu(\vr) = 
    \exp\left(-\frac{y^2+z^2}{w_0^2}\right)\cos \left( \pi \frac{\ell x}{L} \right) \ve_x \;
\end{equation}
of waist $w_0$, mode index $\ell$, and length $L$. (The mode vanishes for $x \notin \left[0,L \right]$, and is laterally well confined within the material.)
The effective volume of the mode is defined as  $V_{\mathrm{eff}} = \int \diff^3 r \, u^2(\vr) = \pi w_0^2 L/4$ and, given the mass density $\varrho$, the effective mass is defined as $m_{\mathrm{eff}} = \int \diff^3 r \, \varrho(\vr) u^2(\vr)$. This way one consistently obtains the total bulk volume and mass in the limiting case of a center of mass motion, corresponding to a uniform displacement with $u(\vr) = 1$, so that $m_{\mathrm{eff}}$ can be understood as the mass fraction taking part in the displacement \cite{Supp}. 

The mode is quantized by means of the ladder operator $\oa = (\ox + i \op)/\sqrt{2}$, involving a conjugate pair of dimensionless position and momentum quadrature operators, $[\ox,\op]=i$. 
A single phononic excitation thus displaces the atoms by amplitudes of the order of the zero-point fluctuation $x_0/\sqrt{2}$, with $x_0 = \sqrt{\hbar/m_{\mathrm{eff}}\omega}$, around their equilibrium positions $ \vr_n$ according to the mode displacement field, \ie $\ovr_n = \vr_n + \vu(\vr_n)x_0 \ox$. The Lindblad operators thus become 
\begin{equation}\label{eq:Lindblad2}
    \oL (\vq) = \frac{1}{m_e} \int\diff^3 r \, \varrho(\vr) e^{i \vq\cdot \vr/\hbar } 
    e^{i\vq\cdot\vu(\vr) x_0 \ox/\hbar } \;.\end{equation}
In the relevant regime of modification length scales $\hbar/\sigma_q$ larger than the interatomic distances, it is safe to assume a homogeneous mass density $\varrho(\vr)=\bar{\varrho}$ and to expand the  operator in \eqref{eq:Lindblad2} to first order in $\ox$ \cite{schrinski2019macroscopicity}. This results in  momentum diffusion, $\cL  \rho = 2 \Gamma \cD[\ox]\rho $, with a diffusion rate amplified by the effective oscillator mass, 
\begin{align}
    \Gamma &= \frac{\bar{\varrho}^2 x_0^2}{2 m_e^2 \tau_e \hbar^2 } \int \diff^3 q\, g(q,\sigma_q)\, q_x^2 \left| \int \diff^3 r\, u_x(\vr) e^{i\vq\cdot\vr/\hbar} \right|^2 \nonumber \\
    &  = \frac{m_{\mathrm{eff}}^2 }{m_e^2 \tau_e} \frac{(4 x_0/L)^2}{1+\sigma_w^2} \int \diff \zeta \, \frac{e^{-\zeta^2/2\sigma_L^2}}{\sqrt{2\pi}\sigma_L} \frac{1-(-)^\ell \cos \zeta}{\left(1 - \pi^2\ell^2/\zeta^2 \right)^2} \; . 
    \label{eq:diffusion}
\end{align}
The rate depends on how the modification length scale $\hbar/\sigma_q$ compares to the geometric length scales of the oscillator mode, as determined by $\sigma_w = w_0\sigma_q/\hbar$, $\sigma_L =  L \sigma_q/ \hbar$, $x_0$, and $\ell$. The integral can be given analytically, leading to a lengthy expression for $\Gamma$.
In the limit $\sigma_L,\sigma_w \ll 1$ and for $\ell$ even, we find $\Gamma \propto m_\mathrm{eff} \sigma_q^2 \sigma_L^4/\ell^4$, whereas for $\sigma_L \gg \pi \ell$ and $\sigma_w \gg 1$, we get $\Gamma \propto m_\mathrm{eff}/\sigma_w^2$. The maximum $\Gamma$ with respect to $\sigma_q$ is located in between at $\hbar/\sigma_q \approx \sqrt{3}L/\pi\ell$, provided that $\pi\ell \gg 1$ and $\pi\ell w_0/\sqrt{3}L \gg 1$ \cite{Supp}, 
\begin{equation}
\max_{\sigma_q}\Gamma \approx \sqrt{\frac{3\pi}{2e^3}} \frac{6 \hbar \bar{\varrho} }{m_e^2 \omega \tau_e} \frac{L}{\ell} \; .   \label{eq:maxGamma} 
\end{equation}
Given a fixed sound velocity $v$, and noting $\omega = 2\pi v \ell/ L$, the maximum rate value is proportional to the square of the mode wavelength $L/\ell$.

In the co-rotating frame of the oscillator, and including energy decay at rate $\gamma_{\downarrow}$,  the modified time evolution of the oscillator state $\rho$ is given by the master equation
\begin{align}
    \dot{\rho} &= \Gamma \cD \left[ \oa e^{-i\omega t} + \oa^\da e^{i\omega t} \right]\rho + \gamma_{\downarrow} \cD [\oa] \rho \nonumber \\
    &\approx \left( \Gamma + \gamma_{\downarrow} \right) \cD[\oa]\rho + \Gamma \cD[\oa^\da]\rho \;.
    \label{eq:modME}
\end{align}
In the second line we average over the rapidly oscillating cross-terms, as we are interested in decoherence on time scales much longer than the oscillation period.

The coarse-grained isotropic master equation \eqref{eq:modME} admits a compact solution in the Wigner function representation
\begin{align}\label{eq:TimeWigner}
    W(X,P;t) = \frac{ e^{\gamma_{\downarrow}t} }{\pi S(t)} \int \diff X' \diff P' & W \left(  X' e^{\gamma_{\downarrow}t/2} ,P' e^{\gamma_{\downarrow}t/2};0  \right)  \nonumber \\
    & \times e^{-[(X-X')^2+(P-P')^2]/S(t)} \;,
\end{align}
with $S(t) = (1 + 2\Gamma/\gamma_{\downarrow})(1-e^{-\gamma_{\downarrow}t})$. This expression can be computed analytically for several states of interest, such as Fock states and their superposition \cite{Supp}. 
{Bayesian parameter estimation based upon Eq.~\eqref{eq:TimeWigner} allows us to identify from the measured Wigner functions the threshold value $\Gamma_{5\%}$ that corresponds to the most conservative $5\,\%$ quantile of the coherence time $\tau_e$ at any $\sigma_q$ \cite{Supp,schrinski2019macroscopicity}. Greater $\tau_e$ are also compatible with the data, as the observed decoherence might as well be due to unspecified conventional noise sources, but smaller times can be ruled out with confidence. 
Macroscopicity is then defined as follows:
We convert $\Gamma_{5\%}$ into the corresponding macrorealistic coherence time parameter $\tau_e = \tau_e (\sigma_q,\Gamma_{5\%})$ by virtue of \eqref{eq:diffusion}, maximize over $\sigma_q$ to obtain the greatest excluded $\tau_e$, and assign a macroscopicity value $\mu=\log_{10}(\tau_e/1\,\mathrm{s})$.}

{By modelling the state evolution with (\ref{eq:modME}), we thus assume a zero-temperature environment and attribute any heating or decoherence beyond energy relaxation at the rate $\gamma_{\downarrow}$ to the impact of a macrorealistic modification described by infinite-temperature diffusion at the rate $\Gamma$. This amounts to a \emph{conservative} assessment of the macroscopicity, since, by overestimating the hypothetical contribution of macrorealistic diffusion to the observed decoherence, we overestimate $\Gamma$ and thus underestimate $\tau_e$.
We characterise the energy relaxation rate $\gamma_{\downarrow}$ independently, by a standard measurement of the phonon decay time $T_1$ and the steady-state population \cite{Supp}.}

\vspace{2mm}
\emph{Experiment.}-- Our measurements are carried out on a high-overtone bulk-acoustic wave resonator (HBAR) device cooled down at millikelvin temperature, see Fig.~\ref{fig:sketch}, which makes use of a superconducting qubit to prepare, control, and read-out the quantum excitation of a mechanical mode localized in the bulk of a sapphire substrate \cite{von2022parity}.

The acoustic vibration mode is described to a very good approximation by Eq.~\eqref{eq:LG00mode}, as also verified by numerical simulations. It oscillates over the bulk length $L=\unit{435}{\mu m}$ at angular frequency $\omega = \unit{2\pi \cdot 5.961}{GHz}$ with wavelength $\lambda=\unit{1.8}{\mu m}$, corresponding to the mode index $\ell=2L/\lambda=486$. 
Assuming a homogenous mass density $\bar{\varrho}=3.98\,$g/cm$^3$, and given the transverse waist $w_0=\unit{27}{\mu m}$, we obtain the effective oscillator mass $m_{\mathrm{eff}} = \pi w_0^2 L \bar{\varrho} /4 = 1.0\,\mu$g and a zero-point fluctuation of $x_0/\sqrt{2}=\unit{1.2 \cdot 10^{-18}}{m}$.
For these parameters, we find from Eq.~\eqref{eq:diffusion} a maximal diffusion rate of $\Gamma_\mathrm{max} = 3.5 \cdot 10^{13}/\tau_e$ at a critical length scale $\hbar/\sigma_q \simeq 0.5\,\mu$m, which is of the same order as the mode wavelength.

The coherence properties of the phonon mode are characterized by a small one-phonon steady-state population of $1.6 \pm 0.2 \,\% $, by the 
relaxation time $T_1 = 85.8{\pm 1.5}\,\mu$s, and by the Ramsey dephasing time $T_2 = 147.3{\pm 2.6}\,\mu$s \cite{Supp}. These imply a much longer pure dephasing time of $T_\phi = (1/T_2-1/2T_1)^{-1} = \unit{1.0{\pm 0.2}}{ms}$.
Relaxation, which originates from the coupling to the environment through \eg surface scattering or diffraction loss, thus dominates the degradation of quantum features over time, and we can safely assume $\gamma_{\downarrow} \approx 1/T_1$ within error tolerance. 

Using the toolbox of circuit quantum acoustodynamics (cQAD), we can use the qubit to prepare and measure the HBAR in nonclassical states of motion \cite{Chu2018}. 
In particular, an arbitrary superposition with the first excited state, $a\ket{0}+b\ket{1}$, can be prepared by swapping the qubit state $a\ket{\downarrow}+b\ket{\uparrow}$ to the mechanical mode through the resonant Jaynes–Cummings interaction. For any $b\neq 0$ we obtain a nonclassical state of motion with negativities in the Wigner function, that will degrade over time due to relaxation. To monitor this effect, we let the system evolve for a time $t$, and then obtain the Wigner function of the mechanical state by measuring its displaced parity~\cite{von2022parity,Supp}. Fig.~\eqref{fig2} shows how the measured Wigner function degrades over time for an initial Fock state $\ket{1}$ (top row) and an initial superposition state $(\ket{0}+\ket{1})/\sqrt{2}$ (bottom row). Note the disappearance of negative (blue) regions in the quasi-probability distribution, turning it into a classical phase-space distribution.

In the first part of the experiment, we prepare the resonator in the single-phonon Fock state $\ket{1}$ and measure its Wigner quasi-probability distribution as a function of time (Fig.~\ref{fig2}, top row). We observe that the region of negative values around the phase space origin shrinks gradually, persisting for tens of microseconds, before the Wigner function turns purely positive and becomes indistinguishable from a classical distribution. The longer and more pronounced the initial negativity remains, the stronger the experiment falsifies macrorealist modifications of quantum mechanics.
We estimate the impact of such modifications by first fitting the initial measurement at $t=0$ with the Wigner function of an incoherent mixture of $\ket{0}$ and $\ket{1}$, thus accounting for imperfections in the state preparation and measurement.
The measured Wigner functions at $t=$ 10\,$\mu$s, 20\,$\mu$s, and 40\,$\mu$s are then compared to the theoretical expectation (\ref{eq:TimeWigner}) including energy relaxation with the measured rate ${\gamma_{\downarrow}\approx} 1/T_1$ as well as modification-induced diffusion with variable rate $\Gamma$. Bayesian parameter estimation and subsequent  maximization with respect to $\sigma_q$ shows that any $\Gamma> \Gamma_{5\%}=\unit{1.6\cdot 10^2}{s^{-1}}$ can be excluded with 95$\%$ confidence (at $\hbar/\sigma_q\simeq \unit{0.5}{\mu m}$) \cite{Supp}.
This corresponds to a macroscopicity of $\mu= 11.3$, by far the highest value reported on mechanical resonators, see Tab.~I. We note that the obtained diffusion rate $\Gamma_{5\%}= \unit{1.6\cdot 10^2}{s^{-1}}$  is consistent with what is found by using our system to perform a so-called non-interferometric test of macrorealistic models based on the observed residual excitation of the mode \cite{Supp}.
Thus our analytic results can be straightforwardly adopted in such schemes, as recently proposed in Ref.~\cite{tobar2022testing}.

In the second part of the experiment, we prepare the resonator in the superposition state $(\ket{0}+\ket{1})/\sqrt{2}$.
As shown in the bottom row of Fig.~\ref{fig2}, the measured Wigner function again takes negative values, confirming the nonclassicality of the initial state. 
The asymmetry around the origin, whose phase can be controlled by the qubit state, rotates due to a small detuning between the oscillator and the displacement pulse frequencies.
This rotation is accounted for in our data processing. Following the previous analysis, we find that the superposition state data is compatible with a maximal  diffusion rate of $\Gamma_{5\%}= \unit{6.4\cdot10^2}{Hz}$, which is a factor of four larger than the rate found from the Fock state $\ket{1}$ and from the non-interferometric test \cite{Supp}. 
This discrepancy can be explained by the presence of an additional pure dephasing channel with a rate of $1/T_\phi\sim \unit{10^3}{Hz}$.
While circularly symmetric Wigner functions such as single Fock states are unaffected by this decoherence channel, which corresponds to a random rotation around the phase-space origin, a superposition of Fock states, which is necessarily asymmetric, gets washed out at rate $1/T_\phi$, removing any negativity. However, to fairly assess the macroscopicity of the Fock state superposition, we do \emph{not} include this decoherence channel into the master equation \eqref{eq:modME}, so that all decoherence is attributed to the macrorealistic modification. This results in a greater diffusion rate, and consequently in a lower macroscopicity of $\mu=10.7$. If we included the $1/T_\phi$ dephasing effect in the Bayesian inference the resulting rate would be consistent with both the Fock state $\ket{1}$ analysis and the non-interferometric test.

\vspace{2mm}
\emph{Beating the macroscopicity record.}--%
Based on our experimental results, and using the analytical result \eqref{eq:diffusion}, we can now establish how to optimize future quantum experiments with bulk acoustic resonators to become competitive with the most macroscopic quantum tests to date.
{In contrast to quantum tests based on the interference of center-of-mass degrees of freedom, the diffusion rate scales here only weakly with the mass and length scales of the material: in the realistic case of comparable lateral and longitudinal mode extension ($w_0 \approx L$), we find that macrorealistic diffusion is strongest at a length scale $\hbar/\sigma_q$ comparable to the mode wavelength $L/\ell$. To attain greater macroscopicity in a given material, one should strive to increase the mode wavelength since the maximum diffusion rate \eqref{eq:maxGamma} grows like $(L/\ell)^2$.}
Apart from these geometric considerations, the macroscopicity also benefits from longer relaxation and coherence times $T_1, T_2$, and a better characterization of the measurement noise. The latter can be incorporated into the Bayesian parameter estimation to extract better bounds, instead of adding a broad Gaussian noise channel as done in the analysis above.        

As an example, keeping the same device and mode geometry as in the present experiment, but using a mode with  $\omega=2\pi\cdot\unit{2}{GHz}$ and $T_1=\unit{10}{ms}$, would result in $\mu=14.4$, surpassing all reported quantum tests, see Tab.~I. 
{We obtain this estimate by setting $\ell=160$ and rescaling the observed coherence time in proportion to $T_1$; the macroscopicity grows logarithmically with increasing coherence time.} 
Beyond that, out-of-plane drum modes with frequencies in the low MHz range and huge quality factors have been reported, corresponding to $T_1>\unit{100}{ms}$ \cite{seis2022ground}. If genuine quantum effects could be demonstrated in these devices, they might reach substantially greater macroscopicities $\mu > 17$ \footnote{This rough estimate is based on \cite{schrinski2019macroscopicity}, by approximating the mode in \cite{seis2022ground} as the fundamental Bessel mode $\vu(\vr)=J_0 (4.81 \sqrt{y^2+z^2}/d) \ve_x$, for a disc with diameter $d=100\,\mu$m, $14\,$nm thickness, and $m_{\mathrm{eff}} \simeq 68\,$pg. Precise $\mu$-values would depend on a combination of effective mass, mode profile, and observed duration of Wigner negativities.}.

\begin{table}[H]
\begin{tabular}{c | l c c}
& experiment & year & $\mu$ \\
\hline
\multirow{3}{5em}{mechanical resonators} & \textbf{Bulk acoustic waves [this work]} & \textbf{2022} & \textbf{11.3}  \\
 & Phononic crystal resonator \cite{Wollack22} & 2022 & $\sim 9.0^*$ \\
& Surface acoustic waves \cite{Satzinger18} & 2018 & $\sim 8.6^*$ \\
\hline
\multirow{3}{5em}{matter-wave interference} & Molecule interferometry \cite{fein2019quantum}  & 2019 & 14.0   \\
& Atom interferometry \cite{xu2019probing} & 2019 & 11.8   \\
& BEC interferometry \cite{asenbaum2017phase}& 2017 & 12.4   \\
\end{tabular}\caption{Macroscopicities of solid-state mechanical resonator experiments reporting Wigner function negativities ($*$: estimated), and most macroscopic matter-wave interference experiments assessed in \cite{schrinski2019macroscopicity,schrinski2020quantum}. The values with asterix are estimates of what could have been observed based on the stated $T_1$ time \cite{Supp}.}
\end{table}

\emph{Conclusion.}--We demonstrated the most macroscopic quantum test with bulk acoustic wave resonators, based on monitoring the time evolution of nonclassical Wigner functions. The reported macroscopicity is comparable to the one obtained by an experiment holding cesium atoms in a spatial superposition at \unit{4}{\mu m} separation over 20 seconds \cite{xu2019probing}. 
In the future, we envisage improved resonator  designs that may surpass the most macroscopic matter wave interferometers, testing the validity of quantum mechanics at unprecedented scales.

%Include the supplementary references here. Use them in Supp entry of bibliography
\nocite{jeffreys1946invariant,ArrangoizPhD,SatzingerPhD,SatzingerPC,ChuSci17,armano2016sub,carlesso2016experimental}

\vspace{5mm}
\emph{Acknowledgements.}--  We would like to thank Georg Enzian and Eric Planz for helpful discussions.  BS was supported by Deutsche Forschungsgemeinschaft (DFG, German Research Foundation), Grant No. 449674892. UvL is funded by the SNF project number 200021\_204073.
MF was supported by The Branco Weiss Fellowship -- Society in Science, administered by the ETH Z\"{u}rich.

\bibliography{letter_PRL_v5}

\clearpage
\newpage

\onecolumngrid
\appendix
\section*{Supplementary information}

\renewcommand\appendixname{Supplementary Section} 
\renewcommand{\thefigure}{S\arabic{figure}}

\section{Geometric factor}\label{app:geometryFactor}

The considered macrorealistic modifications lead to momentum diffusion in the oscillator state. The momentum diffusion rate $D=\hbar^2U/\tau_e$ is inversely proportional to the coherence time parameter $\tau_{ e}$ and depends on the oscillator mode function $\vu(\vr)$ through the geometric factor \cite{schrinski2019macroscopicity}
\begin{equation}\label{eq:UIntGen}
    U=\frac{1}{2\hbar^2}\int d^3q\,  g(\vq,\sigma_q) \vert \tilde{\vu}(\vq) \cdot \vq \vert^2 \;, \qquad \text{with} \qquad \tilde{\vu}(\vq) = \dfrac{1}{m_e} \int d^3r\, \varrho(\vr ) \vu(\vr ) e^{- \vr \cdot\vq / \hbar} \; .
\end{equation}
Here, $m_e$ denotes the reference electron mass, and $g(\vq,\sigma_q) = \exp (-|\vq|^2/2\sigma_q^2)/(2\pi\sigma_q^2)^{3/2}$ a Gaussian momentum distribution of standard deviation $\sigma_q$.
The resonator mode employed in this work can be modelled to great accuracy with the function
\begin{align}\label{eq:DisplacementMode}
    \vu(\vr) = \mathcal{N} e^{-(y^2+z^2)/w_0^2}\cos\left(\pi \dfrac{\ell x}{L} \right)\ve_x \;, 
\end{align}
where $w_0$ is the beam waist ($1/e$ amplitude radius), $L$ is the height, and $\ell$ the mode index. 
Based on this mode function, we can define an effective mode volume and mass, $V_{\mathrm{eff}}=\int \diff^3 r\,|\vu(\vr)|^2$ and $m_{\mathrm{eff}}=\int \diff^3 r\,\varrho(\vr) |\vu(\vr)|^2$. They both depend on the square of the amplitude prefactor $\mathcal{N}$, but the relevant physical displacement amplitudes of the individual atoms, $ x_0 \vu (\vr_n) = \sqrt{\hbar/m_{\mathrm{eff}} \omega} \vu (\vr_n)$, as well as the dimensionless macrorealistic diffusion rate, $\Gamma \tau_e = D (x_0/\hbar)^2 = U x_0^2$, do not depend on $\mathcal{N}$. Hence we may set $\mathcal{N}=1$, without loss of generality, as done in the main text.

Given that we consider a displacement field with only a longitudinal component in $x$-direction, $\vu(\vr)=u_x(\vr)\ve_x$, and that the mass density distribution is assumed to be uniform (\ie $\varrho(\vr )=\bar{\varrho}$), Eq.~\eqref{eq:UIntGen} can be simplified further:
\begin{align}
     U = \frac{1}{2\hbar^2}\int d^3 \vq \, g(\vq,\sigma_q) \vert \tilde{u}_x(\vq) q_x \vert^2
     &= \frac{\bar{\varrho}^2}{2\hbar^2 m_e^2} \int d^3 \vq \, g(\vq,\sigma_q) q_x^2 \left| \int dy\,dz\, e^{-i (q_y y + q_z z )/\hbar - (y^2+z^2)/w_0^2} \int_0^L dx\, \cos \left( \pi \frac{\ell x}{L}\right) e^{-i q_x x /\hbar} \right|^2\nonumber\\
    &=\frac{\bar{\varrho}^2 \pi^2 w_0^4 L^2}{2\hbar^2 m_e^2} \int d^3 \vq \, g(\vq,\sigma_q) q_x^2 e^{-(q_y^2+q_z^2)w_0^2/2\hbar^2} \left| \int_0^1\diff\phi \cos (\pi\ell\phi) e^{-i\phi q_x L/\hbar} \right|^2.
\end{align}
This allows us to first integrate out the momenta $\vq$, change to cylindrical coordinates and express the end result as
\begin{align}
U = \frac{\bar{\varrho}^2}{2m_e^2} \frac{\pi^2 w_0^4}{1+\sigma_w^2} \int \diff \zeta \frac{\zeta^2}{\sqrt{2\pi}\sigma_L} e^{-\zeta^2/2\sigma_L^2} \left| \int_0^1\diff\phi \cos (\pi\ell\phi) e^{-i\phi \zeta} \right|^2 &  = \frac{\bar{\varrho}^2\pi^2 w_0^4}{m_e^2(1+\sigma_w^2)} \int \diff \zeta \, \frac{e^{-\zeta^2/2\sigma_L^2}}{\sqrt{2\pi}\sigma_L} \frac{1-(-)^\ell \cos \zeta}{\left(1 - \frac{\pi^2\ell^2}{\zeta^2} \right)^2}  \; . 
\label{eq:U}
\end{align}
with $\sigma_L= L\sigma_q/\hbar$ and $\sigma_w=w_0\sigma_q /\hbar$.
Once we identify $m_\mathrm{eff} = \bar\varrho \pi w_0^2 L/4$, we arrive at the expression for the diffusion rate $\Gamma = U x_0^2/\tau_e$ given in the main text. 
{A non-tight upper bound follows by pulling the absolute value into the $\phi$-integral in \eqref{eq:U}. The $\phi$-integral then evaluates to $2/\pi$, so that $U < 2 \bar{\varrho}^2 w_0^4 \sigma_L^2/m_e^2(1+\sigma_w^2)$, and $\Gamma < (32/\pi^2)(m_{\mathrm{eff}}^2/m_e^2 \tau_e)(\sigma_q x_0/\hbar)^2/(1+\sigma_w^2)$. The scaling of this bound with the mode mass and geometry is deceivingly similar to previous results for center-of-mass oscillations. However, we will see in the following that the bound is rather loose and the actual diffusion rate for bulk acoustic modes exhibits a completely different scaling.}

{In the limiting case $\sigma_L , \sigma_w \ll 1$, the Gaussian function is sharply peaked around $\zeta = 0$, and we can expand the remainder of the integrand to lowest non-vanishing order; for even $\ell$, this yields $U \approx 15 \bar{\varrho}^2 w_0^4 \sigma_L^6/2\pi^2 m_e^2 \ell^4$, and $\Gamma \approx 120 (m_{\mathrm{eff}}^2/m_e^2 \tau_e)(\sigma_q x_0/\hbar)^2 (\sigma_L/\pi\ell)^4$. For decreasing $\sigma_q$, this rate drops as $\sigma_q^6$.
Notice that odd $\ell$ would result in a $\sigma_q^4$-scaling, $U \approx 3 \bar{\varrho}^2 w_0^4 \sigma_L^4/\pi^2 m_e^2 \ell^4$.}

\begin{figure}[h]
  \centering
\includegraphics[width=0.6\textwidth]{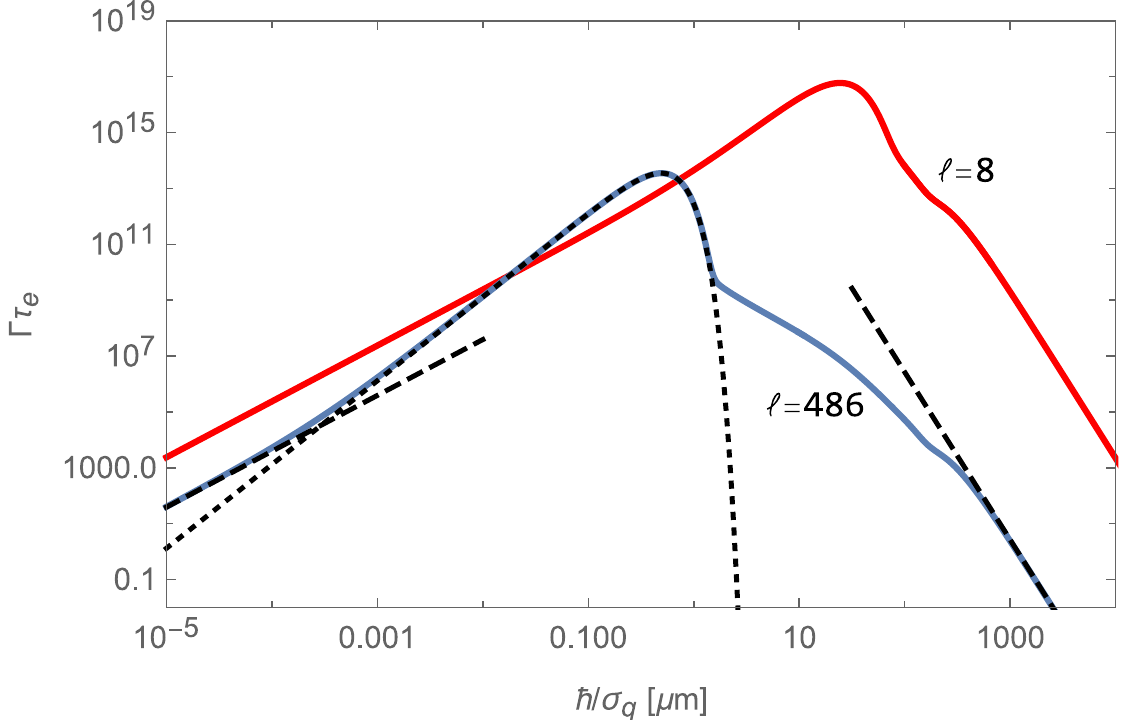}
  \caption{{Dimensionless diffusion rate $\Gamma\tau_e$ as function of the modification parameter $\sigma_q$ for the device studied in the main text (blue curve). We compare it to the correction $U^{(1)}$ from \eqref{eq:U1_approx} (dotted line), as well as to the asymptote $U^{(0)}$ for $\sigma_L \gg \pi\ell$ (left dashed) and to the asymptote for $\sigma_L \ll 1$ (right dashed). Much stronger diffusion would arise (red curve) if instead of the mode at $\ell=486$, we could excite, say, the mode with index $\ell=8$ at the frequency $\omega=2\pi\cdot\unit{98}{MHz}$. }
  }\label{fig:DiffRate}
\end{figure}

{Greater care must be taken in the opposite limit, $\sigma_L, \sigma_w \gg 1$. Although the non-Gaussian part of the integrand in \eqref{eq:U} is clearly peaked around $\zeta = \pm \pi\ell$, it is asymptotically of order $\zeta^0$ and thus non-integrable without the regularizing presence of the Gaussian. Hence a dominant contribution to the $\zeta$-integral in \eqref{eq:U} does not come from the vicinity of the peaks, but from $|\zeta| \gg \pi\ell$. For even greater values $\sigma_L \gg \pi\ell$, we can approximate to lowest order
\begin{align}
    U &\approx U^{(0)} := \frac{\bar{\varrho}^2\pi^2 w_0^4}{m_e^2(1+\sigma_w^2)} \int \diff \zeta \, \frac{e^{-\zeta^2/2\sigma_L^2}}{\sqrt{2\pi}\sigma_L} \left[ 1-(-)^\ell \cos \zeta \right] =  \frac{\bar{\varrho}^2\pi^2 w_0^4}{m_e^2(1+\sigma_w^2)} \left[ 1 - (-)^\ell e^{-\sigma_L^2/2} \right] \approx \frac{\bar{\varrho}^2\pi^2 w_0^4}{m_e^2\sigma_w^2}. 
\end{align}
The corresponding diffusion rate reads as $\Gamma \approx 16 (m_{\mathrm{eff}}^2/m_e^2 \tau_e) (x_0/\sigma_L w_0)^2$, which does not depend on $\ell$, is proportional to $m_{\mathrm{eff}}$, and is again monotonic in $\sigma_q$. In order to access the most relevant intermediate $\sigma_q$-regime of maximum diffusion, consider the correction $U = U^{(0)} + U^{(1)}$, which can be expressed as
\begin{align}
U^{(1)} = \frac{\bar{\varrho}^2\pi^4 \ell^2 w_0^4}{m_e^2(1+\sigma_w^2)} \int \diff \zeta \, \frac{e^{-\zeta^2/2\sigma_L^2}}{\sqrt{2\pi}\sigma_L} \left[ \frac{2\zeta^2 - \pi^2\ell^2}{2(\zeta+\pi\ell)^2} \sinc^2 \left(\frac{\zeta-\pi\ell}{2} \right) \right] \;.
\label{eq:U1}
\end{align}
The non-Gaussian part of this integrand (square-bracketed) is not only sharply peaked at $\zeta = \pm \pi\ell$, but also asymptotically converges to zero like $\zeta^{-2}$. The integral of this expression evaluates to $\pi/2$. Assuming $\pi\ell \gg 1$, we can now replace $\zeta^2$ in the Gaussian by $\pi^2 \ell^2$, so that
\begin{align}
    U^{(1)} \approx \frac{\bar{\varrho}^2\pi^5 \ell^2 w_0^4}{2 m_e^2(1+\sigma_w^2)} \frac{e^{-\pi^2\ell^2/2\sigma_L^2}}{\sqrt{2\pi}\sigma_L} \approx \frac{\bar{\varrho}^2\pi^5 \ell^2 w_0^4 e^{-\pi^2\ell^2/2\sigma_L^2}}{2 \sqrt{2\pi} m_e^2\sigma_w^2 \sigma_L} \; , \quad U\approx \frac{\bar{\varrho}^2\pi^2 w_0^2 L^2}{m_e^2\sigma_L^2} \left( 1 + \frac{\pi^3 \ell^2  e^{-\pi^2\ell^2/2\sigma_L^2}}{2 \sqrt{2\pi} \sigma_L} \right)= \frac{16m_{\mathrm{eff}}^2}{m_e^2\sigma_L^2 w_0^2} \left( 1 + \frac{\pi^3 \ell^2  e^{-\pi^2\ell^2/2\sigma_L^2}}{2 \sqrt{2\pi} \sigma_L} \right) \label{eq:U1_approx}
\end{align}
Presuming $\sigma_w \gg 1$, the local maximum of $U$ coincides approximately with that of $U^{(1)}$, at $\sigma_L \approx \pi\ell/\sqrt{3}$. It marks the $\sigma_q$-value at which the experiment is most sensitive to macrorealistic diffusion. The maximum rate reads as
\begin{equation}
 \max_{\sigma_q}\Gamma \approx \frac{48 m_\mathrm{eff}^2 x_0^2}{m_e^2 \tau_e \pi^2 \ell^2 w_0^2} \left( 1 + \sqrt{\frac{3\pi}{8e^3}} \pi\ell \right) \stackrel{\pi\ell \gg 1}{\approx} \sqrt{\frac{3\pi}{2e^3}} \frac{24}{\pi\ell \tau_e}\left( \frac{m_\mathrm{eff} x_0}{m_e w_0} \right)^2 = \sqrt{\frac{3\pi}{2e^3}} \frac{6 \hbar \bar{\varrho} }{m_e^2 \omega \tau_e} \frac{L}{\ell} \; .
\end{equation}
For numerical evaluation, an exact expression of the diffusion rate can be given in terms of
\begin{equation}\label{eq:GeoFunc_App}
    U = \frac{\bar{\varrho}^2}{2 m_e^2} \frac{\pi^2 w_0^4 \sigma_L^2}{1+\sigma_w^2} f_\ell(\sigma_L)\;, \qquad \text{with} \qquad f_\ell (\xi) = (1+\xi\partial_\xi)
    \int_{[0,1]^2}\!\!
    \diff \phi_1 \diff \phi_2 \, \cos \ell\pi\phi_1 \cos \ell\pi\phi_2 \, e^{- \xi^2 (\phi_1-\phi_2)^2/2} \;.
\end{equation}
}
The function $f_\ell$ can be evaluated to
\begin{align}
    f_\ell (\sigma_L) = \frac{1}{\sqrt{2}\sigma_L^5}\left\{-\ell\pi h(\sigma_L,0,\ell)+\sqrt{2}\sigma_L\left[(-)^\ell e^{-\sigma_L^2/2}-1\right]\left(\ell^2\pi^2-2\sigma_L^2\right)+(-)^\ell \ell\pi^2\mathrm{Re}\left[h(\sigma_L,\sigma_L,\ell)\right]\right\} \;,
\end{align}
where
\begin{align}
    h(a,b,\ell)=-i\frac{\sqrt{\pi}}{2}[6a^2-2\ell\pi(\ell\pi-ib^2) ]\mathrm{erf}\left[\frac{i(\ell\pi/a-ib)}{\sqrt{2}}\right]\exp\left[-\frac{(\ell\pi/a-ib)^2}{2}\right] \;.
\end{align}
In Fig.~\ref{fig:DiffRate}, we plot the resulting  diffusion rate $\Gamma=D/\hbar m_\mathrm{eff}\omega=\hbar U/ m_\mathrm{eff}\omega\tau_e$ as a function of the modification parameter $\sigma_q$. We compare the mode addressed in our experiment to one with a much lower frequency and mode index. 
{Based on the above reasoning, the maxima are expected to be at $\hbar/\sigma_q \approx 0.5\,\mu$m ($\ell=486$) and $17\,\mu$m ($\ell=8$), which roughly agrees with the exact positions on the logarithmic scale. The dotted line represents $U^{(1)}$ from \eqref{eq:U1_approx}, a good approximation in the relevant regime around maximum diffusion. The dashed lines show the asymptotic approximations for $\sigma_L \gg \pi\ell$ (left) and for $\sigma_L \ll 1$ (right). }

\newpage
\section{Initial states and time evolution} \label{app:WignerSolution}

In our setup, macrorealistic collapse models are put to the test by monitoring the decay of Fock and superposition states over a finite amount of time. To assess the influence of those models, it is convenient to solve the time evolution in the Wigner phase space representation of the oscillator state. The one-dimensional Wigner function is defined as
\begin{align}\label{eqapp:WignerFunction}
W(x,p)=\frac{1}{2\pi \hbar}\int \diff s\,e^{ips/\hbar}\langle x-s/2|\rho|x+s/2\rangle \;,
\end{align}
where $x$ and $p$ are the position and momentum coordinates of the harmonic oscillator. For better readability we now change to dimensionless units $X=\sqrt{m_\mathrm{eff}\omega/\hbar}x$ and $P=\sqrt{1/\hbar m_\mathrm{eff}\omega}p$. Then the Wigner functions \eqref{eqapp:WignerFunction} of the ground state $|0\rangle$, first exited Fock state $|1\rangle$, and the superposition state $(|0\rangle+|1\rangle)/\sqrt{2}$, respectively, read as
\begin{align}\label{eq:WignerExperiment}
W_{0}(X,P)=\frac{1}{\pi}e^{-X^2-P^2},\quad
W_{1}(X,P)=\frac{1}{\pi}\left[2X^2+2P^2-1\right]e^{-X^2-P^2},
\quad\mathrm{and}\quad
W_{10}(X,P)=\frac{1}{2\pi}\left[(1+\sqrt{2}X)^2-1+2P^2\right]e^{-X^2-P^2}.
\end{align}

The time evolution of the state described by the master equation \eqref{eq:modME} in the main text describes the impact of the macrorealistic diffusion at the rate $\Gamma$, averaged over the free oscillation, and it includes also an environmental decay channel at the rate $\gamma_{\downarrow}$.
The two separate Lindblad generators for decay and amplification, $\partial_t^\mathrm{dec}\rho = \Gamma \cD[\oa]\rho$ and $\partial_t^\mathrm{amp}\rho = \Gamma \cD[\oa^\da]\rho$ with $\oa=(X+iP)/\sqrt{2}$, translate into phase space as
\begin{align}
\partial_t^\mathrm{dec} W(X,P) &= \frac{\Gamma}{2\pi}\int \diff s\,e^{iP s}\langle X-s/2|\oa\rho\oa^\dagger-\frac{1}{2}\{\oa^\dagger\oa,\rho\}|X+s/2\rangle
=\frac{\Gamma}{4}\left[\partial_{X}^2+\partial_{P}^2+2\partial_{X} X+2\partial_{P} P\right]W(X,P) \;, \\
\partial_t^\mathrm{amp} W(X,P) &= \frac{\Gamma}{4}\left[\partial_{X}^2+\partial_{P}^2-2\partial_{X} X-2\partial_{P} P\right]W(X,P) \;.
\end{align}
Both are classical Fokker-Planck equations, and the first one describes thermalization at the (dimensionless) energy $1/2$, \ie the ground state energy of the quantum oscillator. The master equation then yields 
\begin{align}
 \partial_t W(X,P) = \frac{\Gamma}{2}\left[\partial_{X}^2+\partial_{P}^2 \right] + \frac{{\gamma_{\downarrow}}}{4} \left[\partial_{X}^2+\partial_{P}^2 + 2\partial_{X} X + 2\partial_{P} P\right]W(X,P).    
\end{align}
The solution to this Fokker-Planck equation is obtained with help of a Fourier transform into the characteristic function representation,
\begin{align}
\chi(x,p)=\int \diff x' \diff p'\, W(x',p') e^{i(xp'-px')/\hbar} \;,
\end{align}
which evolves according to
\begin{align}
\partial_t \chi(X,P)
=\frac{1}{2}\left[-\left(\frac{{\gamma_{\downarrow}}}{2}+\Gamma\right)(X^2+P^2)-{\gamma_{\downarrow}}( X\partial_{X}+ P\partial_{P})\right]\chi(X,P) \;.
\end{align}
We can now separate the simple relaxation dynamics (an exponential contraction toward the origin) from the diffusion. Let $X=\tilde{X} e^{{\gamma_{\downarrow}} t/2}$, $P=\tilde{P} e^{{\gamma_{\downarrow}} t/2}$, and $\Phi (\tilde{X},\tilde{P};t) = \chi (\tilde{X}e^{{\gamma_{\downarrow}} t/2},\tilde{P}e^{{\gamma_{\downarrow}} t/2};t)$, which then obeys
\begin{align}
    \partial_t \Phi(\tilde{X},\tilde{P};t) &=  - \frac{{\gamma_{\downarrow}} + 2\Gamma}{4} e^{{\gamma_{\downarrow}} t} (\tilde{X}^2+\tilde{P}^2)\Phi(\tilde{X},\tilde{P};t) \;, \\
\intertext{as solved by} 
\Phi(\tilde{X},\tilde{P};t) &= \exp \left[ - \left(1+\frac{2\Gamma}{{\gamma_{\downarrow}}}\right) (e^{{\gamma_{\downarrow}} t}-1) \frac{\tilde{X}^2+\tilde{P}^2}{4} \right] \Phi(\tilde{X},\tilde{P};0)  \;,
\intertext{so that}
     \quad \chi(X, P;t) &= \exp \left[ - \left(1+\frac{2\Gamma}{{\gamma_{\downarrow}}}\right) (1-e^{-{\gamma_{\downarrow}} t}) \frac{X^2+P^2}{4} \right]  \chi(X e^{-{\gamma_{\downarrow}} t/2}, P e^{-{\gamma_{\downarrow}} t/2};0) \;.
\end{align}
From this expression it is evident that the time evolution of an initial $\chi (X,P;0)$ is simply given by a multiplication with a Gaussian exponential combined with an isotropic rescaling of the arguments. Noting that $\chi(0,0;t) = 1$ at any time, one can check that the asymptotic steady state is a Gaussian, $\chi(X,P;\infty) = \exp [-(1+2\Gamma/{\gamma_{\downarrow}})(X^2 + P^2)/4] $, which for $\Gamma=0$ is the ground state of the harmonic oscillator.

Back in the Wigner representation, the solution turns into a rescaling combined with a Gaussian convolution. Introducing the dimensionless quantitiy $\tilde{T}=1/2+\Gamma/{\gamma_{\downarrow}}$, we get
\begin{align}
    W(X,P;t) &= \frac{1}{4\pi^2} \int \diff \tilde{X} \diff \tilde{P} \, e^{i\tilde{P}X-iP\tilde{X}} \exp \left[ - \tilde{T} (1-e^{-{\gamma_{\downarrow}} t}) \frac{\tilde{X}^2+\tilde{P}^2}{2} \right]  \chi(\tilde{X} e^{-{\gamma_{\downarrow}} t/2}, \tilde{P} e^{-{\gamma_{\downarrow}} t/2};0) \nonumber \\
    &= \frac{1}{4\pi^2} \int \diff \tilde{X} \diff \tilde{P} \diff X_0 \diff P_0 \, W(X_0,P_0;0) \exp \left[ - \tilde{T} (1-e^{-{\gamma_{\downarrow}} t}) \frac{\tilde{X}^2+\tilde{P}^2}{2} + i\tilde{X} (P_0 e^{-{\gamma_{\downarrow}} t/2}-P) - i\tilde{P} (X_0e^{-{\gamma_{\downarrow}} t/2}-X) \right] \nonumber \\
    &= \frac{1}{2\pi \tilde{T} (1-e^{-{\gamma_{\downarrow}} t})} \int \diff X_0 \diff P_0 \, W(X_0,P_0;0) \exp \left[ - \frac{ (X-X_0e^{-{\gamma_{\downarrow}} t/2})^2 + (P-P_0e^{-{\gamma_{\downarrow}} t/2})^2 }{2\tilde{T}(1-e^{-{\gamma_{\downarrow}} t})} \right] \nonumber \\
    &= \frac{e^{{\gamma_{\downarrow}} t}}{2\pi \tilde{T} (1-e^{-{\gamma_{\downarrow}} t})} \int \diff X_0 \diff P_0 \, W(X_0e^{{\gamma_{\downarrow}} t/2},P_0e^{{\gamma_{\downarrow}} t/2};0) \exp \left[ - \frac{ (X-X_0)^2 + (P-P_0)^2 }{2\tilde{T}(1-e^{-{\gamma_{\downarrow}} t})} \right] \;.
\end{align}
The two last lines are alternative ways to write the convolution. Given the simple Gaussian form of the initial Wigner functions \eqref{eq:WignerExperiment}, the integrals can be calculated analytically and result in
\begin{align}
W_0(X,P;t)=&\frac{1}{\pi R(t)}\exp\left[{\gamma_{\downarrow}} t-\frac{e^{{\gamma_{\downarrow}} t}(X^2+P^2)}{R(t)}\right] \;,\\
W_1(X,P;t)=&\frac{4\tilde{T}^2+2e^{{\gamma_{\downarrow}} t}(X^2+P^2+2(e^{{\gamma_{\downarrow}} t}-2)\tilde{T}^2)-1}{\pi R^3(t)}\exp\left[\frac{{\gamma_{\downarrow}} t(1-2\tilde{T})-e^{{\gamma_{\downarrow}} t}(X^2+P^2-2 \tilde{T}{\gamma_{\downarrow}} t)}{R(t)}\right] \;, \label{W1analytical}\\
W_{10}(X,P;t)=&\frac{\sqrt{8}e^{3{\gamma_{\downarrow}} t/2}X\tilde{T}+4e^{2{\gamma_{\downarrow}} t}\tilde{T}^2-\sqrt{2}e^{{\gamma_{\downarrow}} t/2}X(2\tilde{T}-1)+2\tilde{T}(2\tilde{T}-1)+e^{{\gamma_{\downarrow}} t}(P^2+X^2+2\tilde{T}-8\tilde{T}^2)}{\pi R^3(t)}\nonumber\\
&\times\exp\left[\frac{{\gamma_{\downarrow}} t(1-2\tilde{T})-e^{{\gamma_{\downarrow}} t}(X^2+P^2-2 \tilde{T}{\gamma_{\downarrow}} t)}{R(t)}\right] \;, \label{W01analytical}
\end{align}
with $R(t)=1+2(e^{{\gamma_{\downarrow}} t}-1)\tilde{T}$.
These expressions can be directly compared to the experimentally reconstructed Wigner function, which allows us to perform  Bayesian parameter estimation of the macrorealistic $\Gamma$-values that are compatible with the measured data.  

\begin{figure*}
  \centering
\includegraphics[width=16cm]{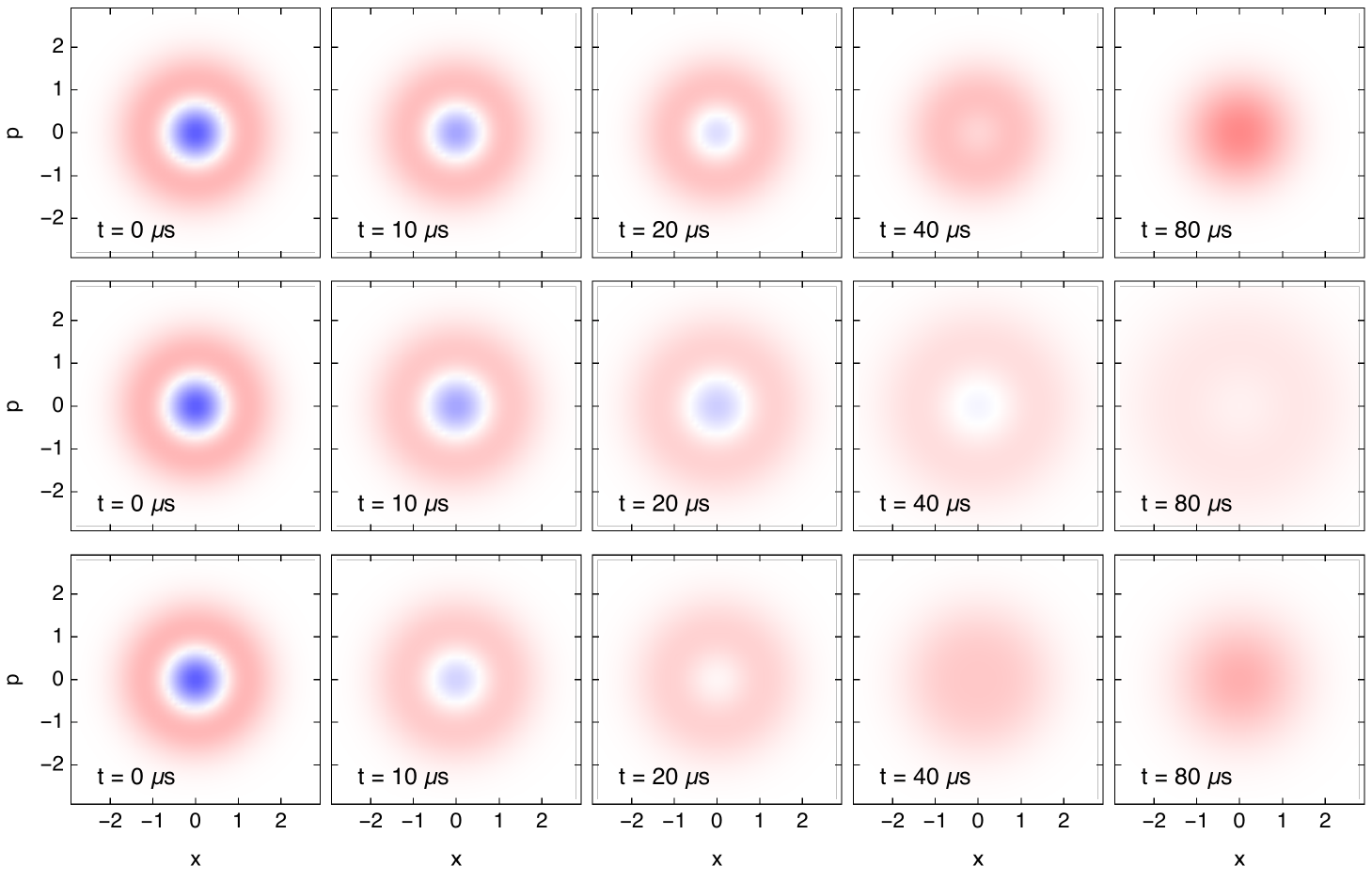}
\includegraphics[width=16cm]{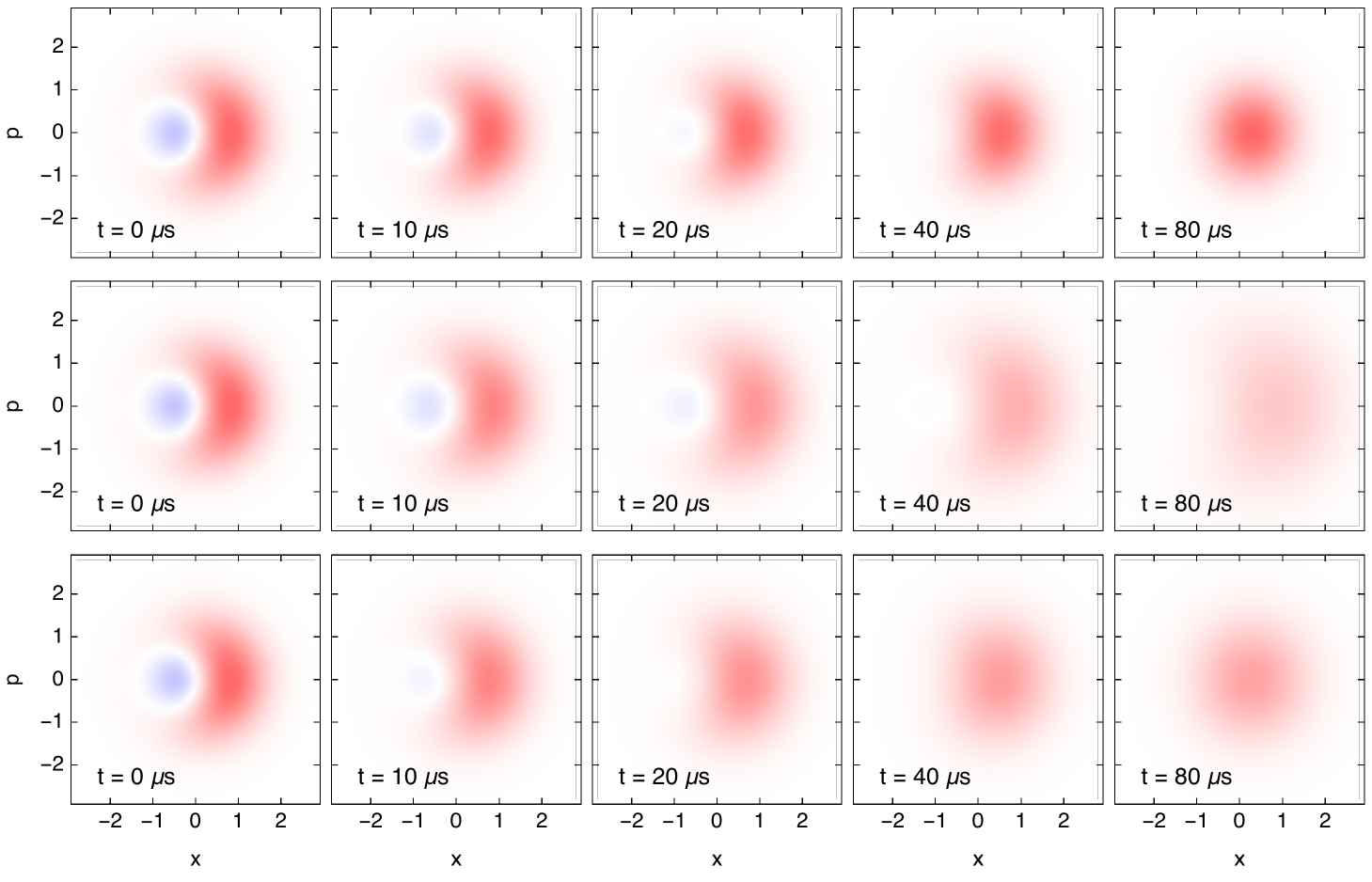}
  \caption{{Comparison between relaxation, diffusion, and a combination of both.} Rows 1 and 4: State $\ket{1}$ and $(\ket{0}+\ket{1})/\sqrt{2}$ evolved according to a relaxation process with $1/\gamma_{\downarrow}=\unit{40}{\mu s}$. Rows 2 and 5: same initial states, but evolved according to a purely dissipative process with $\Gamma=\unit{10^4}{Hz}$. Rows 3 and 6: evolution of the initial states according to a combination of both relaxation and diffusion with same rates. Note that in the actual experiment we have ${\gamma_{\downarrow}}\gg\Gamma$, thus diffusion contributes very little compared to relaxation.
  }\label{figSuppComp}
\end{figure*}

In Fig.~\ref{figSuppComp}, we use Eqs.~(\ref{W1analytical}), (\ref{W01analytical}) to illustrate the evolution of states $\ket{1}$ and $(\ket{0}+\ket{1})/\sqrt{2}$ according to either pure relaxation, pure diffusion, or a combination of both. Notice the  significant differences in the dynamics between the different cases.

\clearpage
\newpage

\section{Bayesian parameter estimation}\label{app:BayesDetails}

\begin{figure*}
  \centering
\includegraphics[width=\textwidth]{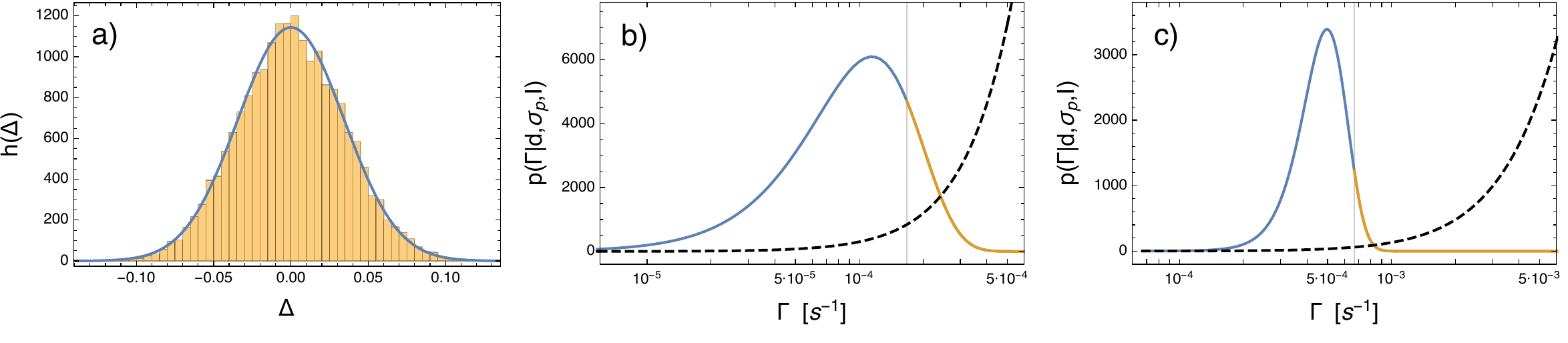}
  \caption{{Bayesian inference of the diffusion rate.} (a): Histogram of pixel deviations $\Delta_{ij} (t)$ between recorded data and model Wigner function, sampled from four time snapshots of the Fock state measurement. The histogram is closely matched by a Gaussian fit (solid line) with standard deviation $s=0.034$. (b): Posterior distribution for the diffusion rate $\Gamma$ for the single Fock state data up at $t=10,20,40\,\mu$s. For comparison, the dashed curve depicts Jeffreys' prior, which has been updated by the posterior toward significantly lower $\Gamma$-values. The use of Jeffreys' prior leads to a more conservative estimate of excluded $\Gamma$-values than, say, the often used flat prior. The vertical line and the yellow part mark the upper five percent quantile of the posterior. (c) Posterior distribution for the superposition state data.
  }\label{fig:BayesianInference}
\end{figure*}

By measuring the Wigner function at several instances of time and comparing it to the theoretical predictions in Sect.~\ref{app:WignerSolution}, one can infer the most likely values for the macrorealistic diffusion rate $\Gamma$ and, subsequently, for the values of the time parameter $\tau_e= \hbar U/\Gamma m_\mathrm{eff}\omega$ by virtue of Bayes' rule. The posterior distribution of $\tau_e$-values given the measured data $d$, a fixed value of the parameter $\sigma_q$, and an independently inferred decay time ${1/\gamma_{\downarrow}\approx T_1}$, updates as \cite{schrinski2019macroscopicity}
\begin{align}\label{eq:BayesianInference}
p(\tau_e|d;\sigma_q,T_1)\propto p(d|\tau_e;\sigma_q,T_1)p(\tau_e|\sigma_q,T_1), \end{align}
Here, $d = [d_{ij} (t)]$ subsumes all recorded pixel values $d_{ij} (t) \in \mathbb{R}$ of the two-dimensional Wigner function reconstructions at the times $t=10,20,\unit{40}{\mu s}$, as plotted in the main text.
The likelihood $p(d|\tau_e;\sigma_q,T_1)$ for $d$ follows from the theoretical model for the Wigner function evaluated at the recorded pixel coordinates and times, $W(X_i,P_j;t)$, and the assumption of additional Gaussian noise in each data point. That is, we model the likelihood for each pixel value by a Gaussian distribution of standard deviation $s$, so that 
\begin{align}
p(d|\tau_e;\sigma_q,T_1) = \prod_{i,j,t} \frac{1}{\sqrt{2\pi s^2}}\exp \left\{- \frac{\left[d_{ij} (t) -W(X_i,P_j;t) \right]^2}{2 s^2} \right\}\;.  
\end{align}
We extract the overall noise level $s$ by taking a sample of all pixel values for the Fock-$\ket{1}$ measurements at $t=0,10,20,\unit{40}{\mu s}$ and subtracting the respective model Wigner functions $W_1(X,P;t)$ with decay time $T_1$ and without macrorealistic diffusion ($\Gamma=0$). A histogram of the resulting pixel deviations $\Delta_{ij} (t) = d_{ij} (t) - W_1(X_i,P_j;t)$ is shown in Fig.~\ref{fig:BayesianInference}(a), together with a Gaussian fit that yields the noise level $s=0.034$.

The prior $p(\tau_e|\sigma_q,T_1)$ is chosen as Jeffreys' prior \cite{jeffreys1946invariant}, which is maximally objective and fair in a comparison with other completely different experiments \cite{schrinski2019macroscopicity}. It is defined as
\begin{align}
p(\tau_e|\sigma_q,T_1)\propto\sqrt{\cI(\tau_e|\sigma_q,T_1)}=
\sqrt{\left\langle\left(\frac{\partial}{\partial\tau_e}\log p(d|\tau_e;\sigma_q,T_1)\right)^2\right\rangle_d},
\end{align}
where $\cI(\tau_e|\sigma_q,I)$ is the Fisher information of the likelihood with respect to $\tau_e$ and $\langle\cdot\rangle_d$ denotes the expectation value over all possible measurement results, $d_{ij} (t)\in \mathbb{R}$. 

We perform the Bayesian update \eqref{eq:BayesianInference} with the data obtained for the three time snapshots at $10,20,\unit{40}{\mu s}$. Figures \ref{fig:BayesianInference}(b) and (c) show the resulting posterior distributions over $\Gamma$-values for the Fock-$\ket{1}$ and the superposition state, respectively. The upper five-percent quantiles for $\Gamma$, marked by vertical lines, correspond to lower five-percent quantiles of $\tau_e= \hbar U/\Gamma m_\mathrm{eff}\omega$ at a given value for $\sigma_q$. Smaller time parameters are thus ruled out with 95\% confidence. We find the threshold values of $\Gamma=\unit{1.6\cdot 10^2}{s^{-1}}$ and $\unit{6.4\cdot 10^2}{s^{-1}}$ in (b) and (c), respectively. More conservatively, the respective bounds would be $\Gamma=\unit{3.1\cdot 10^2}{s^{-1}}$ and $\unit{8.3\cdot 10^2}{s^{-1}}$ at the confidence level $1-10^{-3}$ (roughly corresponding to $3\sigma$ for normally distributed estimates), or $\Gamma=\unit{5.5\cdot 10^2}{s^{-1}}$ and $\unit{1.1\cdot 10^3}{s^{-1}}$ at the confidence level $1-10^{-7}$ (roughly $5\sigma$). 
We attribute the greater threshold value in the superposition measurement to the omitted influence of environmental dephasing, which occurs at a comparable rate of $1/T_\phi=1.0\cdot\unit{10^3}{s^{-1}}$, as discussed in the main text. The remaining discrepancy between this rate and the inferred $\Gamma$ can be explained by the different impact of diffusion and pure dephasing on the oscillator state.

In order to obtain the macroscopicity values, we take the maximum of the ruled out $\tau_e$-range over $\sigma_q$, which coincides with the local maximum of the dimensionless diffusion rate at $\hbar/\sigma_q = \unit{0.5}{\mu m}$, as depicted in Fig.~\ref{fig:DiffRate}. Employing the same 95\% confidence level as in previous assessments, this leads to the macroscopicity values $\mu=11.3$ and $10.7$ for the two measured states.

\newpage
\section{Macroscopicity estimates for other mechanical resonators}

We report here the parameters used to estimate the macroscopicities reachable in the two other reported experiments with mechanical resonators, where Wigner function negativities have been measured in the initial state \cite{Satzinger18,Wollack22}. These experiments involve resonators with acoustic modes of complicated geometry \cite{ArrangoizPhD,SatzingerPhD}, so that it is not easy to derive simple analytical expressions as for the Laguerre-Gaussian modes we considered. For this reason, we proceed with making a conservative analysis, based on approximating such devices to cuboids hosting a sinusoidal displacement field, $\vu (\vr) = \cos \left( \pi/h \right) \ve_x$ with $h$ the longitudinal extension of the mode. This yields an analytic expression for the geometric factor \eqref{eq:UIntGen} determining macrorealistic diffusion, which would otherwise require numerical integration of the displacement field resulting from a finite-element simulation of the actual device. Our approximation is conservative in the sense that it likely overestimates the effective oscillator mass, but we do not expect it to result in macroscopicities that deviate much from the actual one (given that $\mu$ is in logarithmic scale).

The experiments in Refs.~\cite{Satzinger18,Wollack22} show measurements of Wigner function negativities with the sake of demonstrating nonclassical state preparation, but they do not monitor the disappearance of such quantum features with time. 
We can therefore merely estimate the potential of these setups to perform a test of macrorealist models from the limitations imposed by the reported $T_1$-times. To this end, we simply assume that our data were obtained with their resonators at the same \emph{relative} times $t/T_1$, which amounts to rescaling our inferred $\Gamma$-values by the ratio between our $T_1$ time and theirs.

\textbf{Phononic crystal resonator Ref.~\cite{Wollack22}:} 
We consider a phonon mode with frequency $\omega=\unit{2\pi\cdot 2}{GHz}$, wavelength $\lambda=\unit{1}{\mu m}$, and decay time $T_1=\unit{1}{\mu s}$.  The resonator itself is approximated by a lithium niobate (density $\unit{4.65}{g \, cm^{-3}}$) cuboid of size $1\times 1\times 0.25 \,\unit{}{\mu m}$ and total mass $m=\unit{1.16 \cdot 10^{-15}}{kg}$, with an effective mass of $m_{\text{eff}}=5.8\cdot\unit{10^{-16}}{kg}$. We assume a hypothetical experiment that excludes diffusion rates $\Gamma > \unit{13.7}{kHz}$, which results in $\mu = 9.0$ after maximizing with respect to $\sigma_q$.

\textbf{Surface acoustic waves Ref.~\cite{Satzinger18}:} 
We consider a phonon mode with frequency $\omega=\unit{2\pi\cdot 4}{GHz}$, wavelength $\lambda=\unit{1}{\mu m}$, and decay time $T_1=\unit{150}{n s}$. The resonator is approximated by a lithium niobate (density $\unit{4.65}{g\,cm^{-3}}$) cuboid of size \cite{SatzingerPC} $75\times 50\times 1 \,\unit{}{\mu m}$, with total mass $m=\unit{1.744 \cdot 10^{-11}}{kg}$ and effective mass $m_{\text{eff}}=8.7\cdot\unit{10^{-12}}{kg}$. The hypothetical experiment would exclude $\Gamma > \unit{91.5}{kHz}$, resulting in $\mu = 8.6$.

\section{{Experimental sequences}} \label{app:modeDetails}

\begin{figure*}
  \centering
\includegraphics[width=\textwidth]{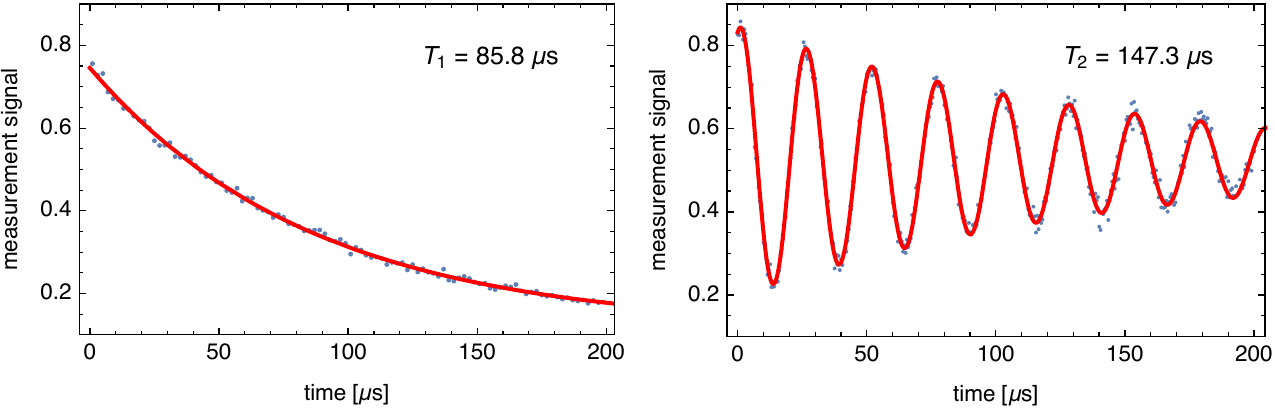}
  \caption{{Characterization of phonon coherence.} Left: population decay measurement used to estimate the relaxation time $T_1$. Right: phonon Ramsey measurement used to estimate the dephasing time $T_2$.
  }\label{figT1T2}
\end{figure*}

\subsection{{Phonon mode characterization}}

To characterize the phonon mode under investigation, we perform two standard measurements \cite{ChuSci17}.
The first is a population decay (``ring-down'') measurement, which gives the relaxation rate $1/T_1$. This is done by preparing the phonon mode in Fock state $\ket{1}$, and then monitoring its population decay over time, see Fig.~\eqref{figT1T2}. Fitting the measured data with a decaying exponential $e^{-t/T_1}$ plus a constant background
gives us the time parameter $T_1=\unit{85.8{\pm 1.5}}{\mu s}$, where the error is one standard deviation given by the fitting function. 
In the assumed model \eqref{eq:modME}, the measured relaxation rate corresponds to $1/T_1 = 2\Gamma + \gamma_{\downarrow}$.

The second measurement consists of a Ramsey sequence, which monitors dephasing effects. This is done by preparing the phonon mode in state $(\ket{0}+\ket{1})/2$, and then monitoring its phase evolution over time, see Fig.~\eqref{figT1T2}.
Fitting the measured data to an exponentially decaying oscillation gives us the time constant $T_2=\unit{147.3{\pm 2.6}}{\mu s}$, where the error is one standard deviation given by the fitting function. As this contrast reduction is partially due to the relaxation effect previously investigated, it is convenient to identify the pure dephasing time as $T_\phi = (1/T_2 - 1/(2T_1))^{-1}=\unit{1.0{\pm 0.2}}{ms}$. This pure dephasing could be described by a Lindblad generator of the form $2 \cD[\oa^\da \oa]/T_\phi$.

\begin{figure*}
  \centering
\includegraphics[width=15cm]{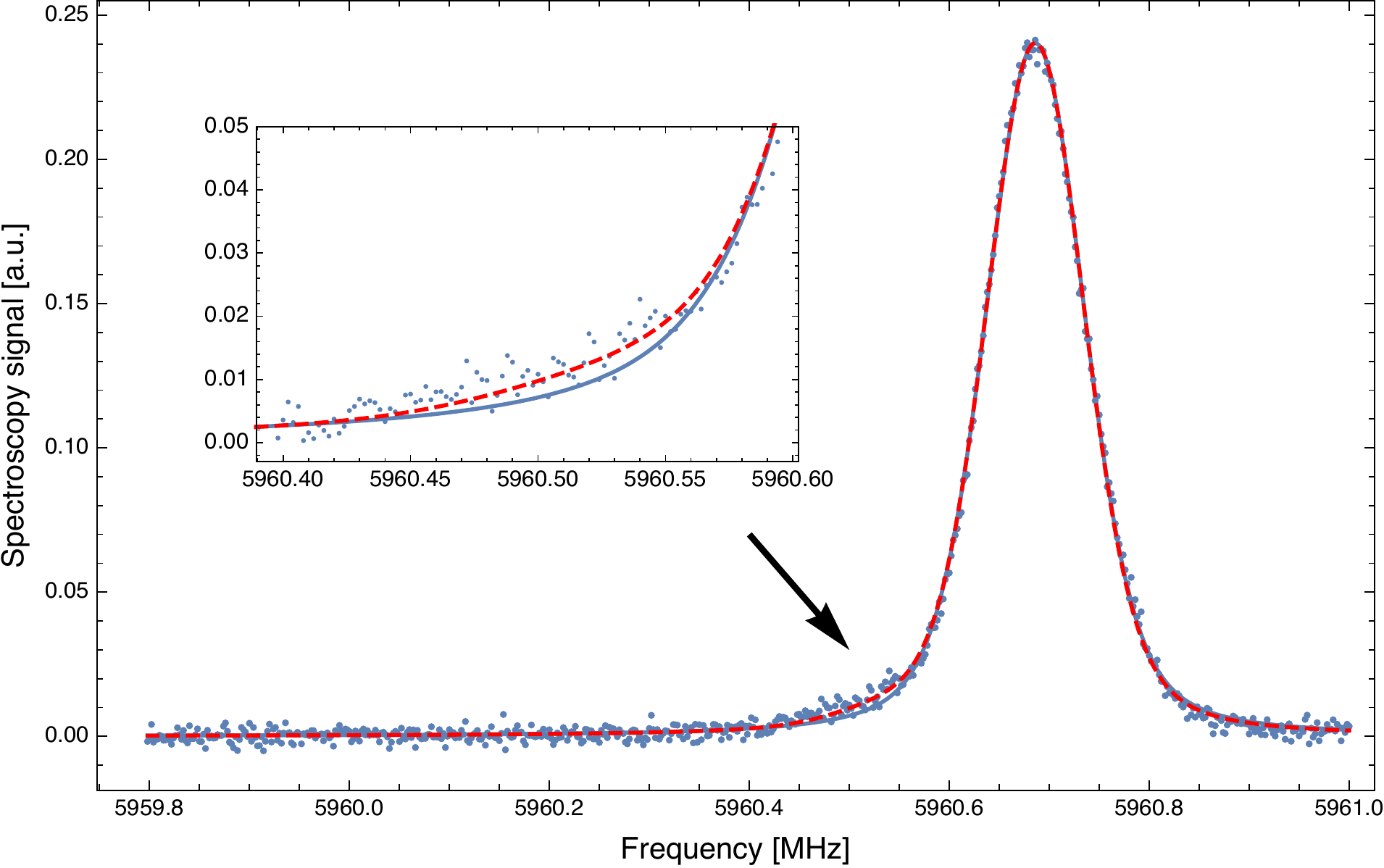}
  \caption{{Phonon mode thermometry.} Spectroscopy of the qubit resonance frequency, while coupled to the phonon mode in the strong dispersive regime. In this regime, the qubit frequency depends on the phonon occupation number. Observing an asymmetry in the measured curve thus indicates a non-zero phonon number, that we associate to an effective temperature (see text for details). 
  }\label{figTh}
\end{figure*}

We can also estimate the ground state population and the effective temperature of the phonon mode, which can be used for a
{classical noise test} 
of macrorealistic collapse models. To this end, we tune the qubit frequency to be \unit{0.8}{MHz} below the one of the phonon mode, operating our system in the strongly dispersive regime \cite{von2022parity}. In this regime, the qubit frequency depends on the phonon mode occupation number, and therefore on its effective temperature. In Fig.~\ref{figTh} we present a spectroscopic measurement of the qubit frequency with the phonon mode in the stationary state, showing a prominent peak at the bare qubit frequency. From the data it is possible to note a slight asymmetry between the left and right tails of the peak, which is highlighted by comparing to a Voigt profile fit (blue line). This is expected at finite temperature, as any population in higher Fock states will result in additional peaks on the left-hand side of the main one. As the position of the peak for $\ket{1}$ is known, we fit again the data with a sum of two Voigt profiles of fixed frequencies (red dashed line). From the relative amplitude between the two peaks we estimate the thermal population of the 1-phonon excited state to be $1.6\pm 0.2\,\%$. In our model \eqref{eq:modME} in the main text, this small population corresponds to $\Gamma/(2\Gamma+\gamma_{\downarrow})$ to a good approximation. Hence $\Gamma \ll \gamma_{\downarrow}$, and we can safely identify the measured relaxation rate with the model rate, $\gamma_{\downarrow} \approx 1/T_1$, within the uncertainty of the fit.

\subsection{{State preparation and read-out}}
\begin{figure*}
  \centering
\includegraphics[width=15cm]{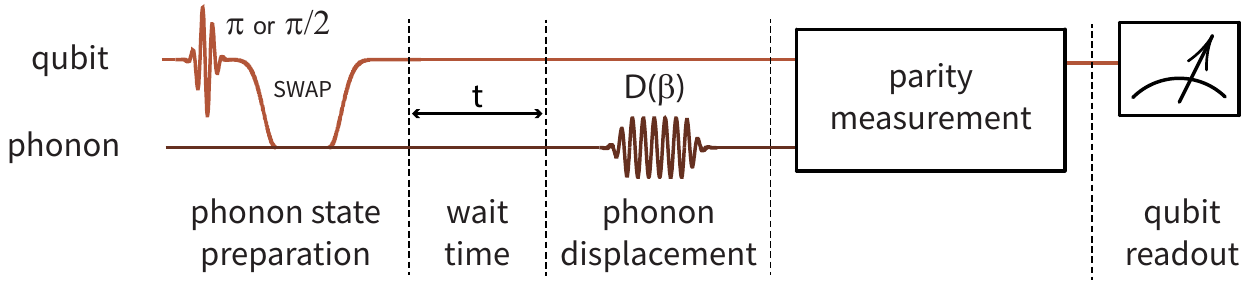}
  \caption{{{Wigner function measurement of the time-evolved state.} The experimental sequence used to record the data in Fig.~\ref{fig2} consists of the following steps: 1) phonon state preparation, 2) wait time $t$, and 3) displaced parity measurement. Further details about the parity measurement sequence are given in Ref.~\cite{von2022parity}.}
  }\label{fig:decSequence}
\end{figure*}

{To acquire the data presented in Fig.~\ref{fig2} of the main text, we adopt the sequence illustrated in Fig.~\ref{fig:decSequence}. First, the phonon mode is prepared in state $\ket{1}$ or $(\ket{0}+\ket{1})/\sqrt{2}$ by swapping the corresponding qubit states $\ket{\uparrow}$ or $(\ket{\downarrow}+\ket{\uparrow})/\sqrt{2}$ with the phonon ground state through the resonant Jaynes–Cummings interaction. Then, the system  evolves freely for a variable time interval $t$, after which a Wigner function measurement is performed. The latter consists of a displaced parity measurement, where a displacement of the phonon mode $D(\beta)$ is followed by a parity measurement implemented through a strong dispersive interaction between the qubit and the phonon~\cite{von2022parity}.}

{The above sequence results in a single measurement of the parity at point $-\beta$ in phase space. The value of the Wigner function is then obtained by averaging $\sim 4 \cdot 10^{3}$ of such parity measurements. Finally, to record a two-dimensional Wigner function like the ones in Fig.~\ref{fig2}, we repeat the described procedure for many different values of $\beta$.}

\newpage
\section{Non-interferometric thermalization test}

A straightforward and widely used approach to study collapse models are so-called non-interferometric tests, which focus on monitoring the noise or the energy changes in the studied system \cite{CarlessoNat22}. This puts bounds on modifications to quantum mechanics without the need of preparing a quantum state. In our experiment, we can do a similar analysis by monitoring the resonator occupation number. The Wigner function evolution under relaxation at the rate ${\gamma_{\downarrow}}$ and macrorealistic diffusion predicts a steady-state energy $E_\mathrm{therm}=\hbar\omega (1+2\Gamma/{\gamma_{\downarrow}})/2$. Hence, by measuring the steady state phonon population, we impose bounds on the hypothetical diffusion rate $\Gamma$. 

To improve the bounds resulting from this analysis, one could also take into account additional known decoherence processes and a finite temperature characterized from independent measurements. Instead, we follow the most conservative estimation: we attribute all diffusion effects to the macrorealistic modification, thus overestimating its effect and underestimating the resulting bounds. Then the equilibrium temperature $T_\mathrm{therm}=\hbar\omega\Gamma/{\gamma_{\downarrow}} k_B $  is a result of the competing relaxation rate ${\gamma_{\downarrow}}$ and macrorealistic diffusion rate $\Gamma$. The decay rate ${\gamma_{\downarrow}}$ can be reliably measured far from equilibrium by preparing a thermal state at $T(t=0)\gg T(t=\infty)=T_\mathrm{therm}$ and verifying a constant decay rate. This leaves only one free parameter $\Gamma$ to estimate, which then gives non-interferometric exclusion bounds on collapse models based on the observed amount of classical heating.

With the maximum dimensionless diffusion rate $\Gamma_\mathrm{max}=3.5\cdot10^{13}/\tau_e$ and a thermal population of $1.6\%$, this analysis leads to $\Gamma=0.016{\gamma_{\downarrow}}$. The measured $ \gamma_{\downarrow} \approx 1/T_1 $ then results in an exclusion of $\tau_e<1.9\cdot 10^{11}\,$s at the length scale $\hbar/\sigma_q = 5\cdot10^{-7}\,$m. This is weaker than other non-interferometric collapse tests performed with \emph{ad hoc} experiments \cite{CarlessoNat22}, or with large-scale endeavors like the LISA pathfinder experiment \cite{armano2016sub,carlesso2016experimental}, but it could be further improved by carefully optimizing the HBAR device geometry and the experimental sequences. 
Note however that no genuine quantum feature is actually verified in a collapse test based on classical noise bounds, meaning that it does not qualify for a benchmark based on the macroscopicity of a superposition.

\begin{figure*}
  \centering
\includegraphics[width=10cm]{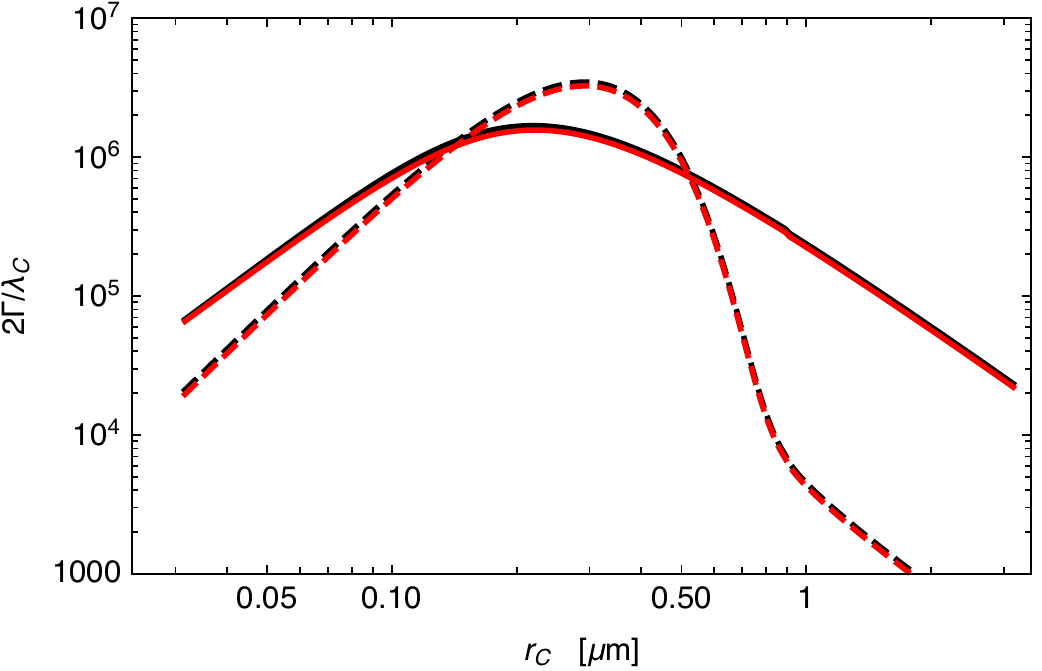}
  \caption{{Comparison between diffusion rates.} Our results (red), compared to the results of Ref.~\cite{tobar2022testing} (black). Solid curve is for a mode index of $\ell=1$, while the dashed curve is for $\ell=40$. We chose to plot $2\Gamma=2\eta x_0^2$ for a direct  comparison. 
  }\label{figBowenResults}
\end{figure*}

Finally, as a side remark, let us mention that a new approach for performing non-interferometric tests of collapse models using bulk resonators has been recently proposed in Ref.~\cite{tobar2022testing}. Our analytical results can readily be applied to this proposal, leading to an exact expression of the diffusion rate. Given the mode function $u(\vr) = \cos ( \pi \ell x/L ) \ve_x$ hosted by a homogeneous cylinder of radius $R$ and length $L$, we obtain the diffusion rate 
\begin{align}
\Gamma=&\lambda_\mathrm{C}\frac{x_0^2\bar{\varrho}^2 \pi^2 r_\mathrm{C}^5 R^2}{2m_e^2L^3}
\left[\frac{\exp(R^2/2r_\mathrm{C}^2)-I_0(R^2/2r_\mathrm{C})-I_1(R^2/2r_\mathrm{C})}{\exp(R^2/2r_\mathrm{C}^2)}\right] \nonumber\\
&\times\left[-\ell\pi h(L/\sqrt{2}r_\mathrm{C},0,n)+\sqrt{2}\left((-1)^{n}e^{-L^2/4r_\mathrm{C}^2}-1\right)\sigma_L\left(\ell^2\pi^2-L^2/r_\mathrm{C}^2\right)+(-1)^n\ell\pi^{2}\mathrm{Re}\left\{h(L/\sqrt{2}r_\mathrm{C},L/\sqrt{2}r_\mathrm{C},\ell)\right\}\right] \;.
\end{align}
For convenience, we have adopted the parameter notation of the CSL collapse model here, $\lambda_\mathrm{C}=m_0^2/m_e^2\tau_e$ and $r_\mathrm{C}=\hbar/\sqrt{2}\sigma_q$. Our formula must be multiplied by two for direct comparison with the approximate momentum diffusion rate term given by the authors,
\begin{align}
2\Gamma \approx &\lambda_\mathrm{C}\frac{x_0^2r_\mathrm{C}^3\bar{\varrho}^2R^2}{2m_0^2}\left(
\frac{64\sqrt{\pi}}{L/\ell}\right)\left[\frac{\exp(R^2/2r_\mathrm{C}^2)-I_0(R^2/2r_\mathrm{C})-I_1(R^2/2r_\mathrm{C})}{2\exp(R^2/2r_\mathrm{C}^2)}\right]  \nonumber\\
&\times\int_{-\infty}^\infty\mathrm{d}a\,\exp\left(-r_\mathrm{C}^2\left(\frac{2\pi a}{L/\ell}\right)^2\right)
\frac{(-8+(8+a^2\pi^2)\cos(a\pi/2))^2}{4a^2\pi^2}
\frac{\sin^2(\ell\pi a/2)}{\cos^2(\pi a/2)}.
\end{align}
A comparison between both expressions is shown in Fig.~\ref{figBowenResults} for the parameters chosen in Ref.~\cite{tobar2022testing}, $\bar{\varrho}=\unit{3210}{kg\, m^{-3}}$, $R=\unit{35}{\mu m}$, $\omega=2\pi\times\unit{6.33}{Hz}$, and for the two settings at $\ell=1, L=\unit{1.5}{\mu m}$ and at $\ell=40, L=\unit{60}{\mu m}$, showing good agreement.

\end{document}